%% file: a1703astroph.tex
\PassOptionsToPackage{pdfpagelabels=false}{hyperref}
\documentclass[fleqn,usenatbib]{mnras}

\usepackage{mathptmx}

\usepackage[T1]{fontenc}
\usepackage{ae,aecompl}

\usepackage{graphicx}	
\usepackage{amssymb}	

\newcommand{\mincir}{\raise -2.truept\hbox{\rlap{\hbox{$\sim$}}\raise5.truept
\hbox{$<$}\ }}
\newcommand{\magcir}{\raise -2.truept\hbox{\rlap{\hbox{$\sim$}}\raise5.truept
\hbox{$>$}\ }}
\newcommand{\siml}{\raise -2.truept\hbox{\rlap{\hbox{$\sim$}}\raise5.truept
\hbox{$<$}\ }}
\newcommand{\simg}{\raise -2.truept\hbox{\rlap{\hbox{$\sim$}}\raise5.truept
\hbox{$>$}\ }}
\newcommand{\be}{\begin{equation}}
\newcommand{\ee}{\end{equation}}
\newcommand{\ba}{\begin{eqnarray}}

\newcommand{\ea}{\end{eqnarray}}

\newcommand {\ks} {km~s$^{-1} \;$}
\newcommand {\kss} {km~s$^{-1}$}
\newcommand {\m} {$M_{\odot} \;$}
\newcommand {\mm} {$M_{\odot}$}

\newcommand {\mmlr} {$M_{\odot}/L_{r,\odot} \;$}

\newcommand{\degree}{\ensuremath{\mathrm{^\circ}}}
\newcommand{\arcm}{\ensuremath{\mathrm{^\prime}\;}}
\newcommand{\arcs}{\ensuremath{\arcmm\hskip -0.1em\arcmm \;}}
\newcommand{\arcmm}{\ensuremath{\mathrm{^\prime}}}
\newcommand{\arcss}{\ensuremath{\arcmm\hskip -0.1em\arcmm}}

\newcommand{\dotsec}{\,\rlap{\hbox{$\mathrm{^s}$}}{\hbox{$.$}}\,}

\newcommand{\chandra}{{\em Chandra}}

\newcommand\rosat{{\sl ROSAT}}

\title[A spectroscopic survey of Abell 1703]{A spectroscopic survey of Abell 1703: is it a rare relaxed cluster hosting a radio halo or a usual merging system?}

\author[W. Boschin et al.]{W. Boschin,$^{1,2,3}$\thanks{E-mail: boschin@tng.iac.es}
M. Girardi,$^{4,5}$
and F. Gastaldello$^{6}$
\\
$^{1}$Fundaci\'on G. Galilei - INAF (Telescopio Nazionale Galileo),
  Rambla J. A. Fern\'andez P\'erez 7, E-38712 Bre\~na Baja (La Palma),
  Spain\\
$^{2}$Instituto de Astrof\'{\i}sica de Canarias, C/V\'{\i}a L\'actea
  s/n, E-38205 La Laguna (Tenerife), Spain\\
$^{3}$Departamento de Astrof\'{\i}sica, Univ. de La Laguna, Av. del
  Astrof\'{\i}sico Francisco S\'anchez s/n, E-38205 La Laguna
  (Tenerife), Spain\\
$^{4}$Dipartimento di Fisica dell'Universit\`a degli Studi di Trieste
  - Sezione di Astronomia, via Tiepolo 11, I-34143 Trieste, Italy\\
$^{5}$INAF - Osservatorio Astronomico di Trieste, via Tiepolo 11,
  I-34143 Trieste, Italy\\
$^{6}$INAF - IASF Milano, via E. Bassini 15, 20133, Milano, Italy
}

\date{Accepted: 2019 December 23; Revised: 2019 November 27; Received: 2019 October 15}
\pubyear{2020}

\begin{document}
\label{firstpage}
\pagerange{\pageref{firstpage}--\pageref{lastpage}}
\maketitle

\begin{abstract}

We present the study of the internal dynamics of the intriguing galaxy
cluster Abell 1703, a system hosting a probable giant radio halo whose
dynamical status is still controversial. Our analysis is based on
unpublished spectroscopic data acquired at the Italian Telescopio
Nazionale {\it Galileo} and data publicly available in the
literature. We also use photometric data from the Sloan Digital Sky
Survey. We select 147 cluster members and compute the cluster redshift
$\left<z\right>\sim 0.277$ and the global line-of-sight velocity
dispersion $\sigma_{\rm v}\sim 1300$ \kss. We infer that Abell 1703 is
a massive cluster: $M_{200}\sim 1-2\times 10^{15}$\mm. The results of
our study disagree with the picture of an unimodal, relaxed cluster as
suggested by previous studies based on the gravitational lensing
analysis and support the view of a perturbed dynamics proposed by
recent works based on {\it Chandra} X-ray data. The first strong
evidence of a dynamically disturbed cluster comes from the peculiarity
of the BCG velocity with respect to the first moment of the velocity
distribution of member galaxies. Moreover, several statistical tests
employed to study the cluster galaxies kinematics find significant
evidence of substructure, being Abell 1703 composed by at least two or
three subclumps probably caught after the core-core passage. In this
observational scenario, the suspected existence of a radio halo in the
centre of this cluster is not surprising and well agrees with the
theoretical models describing diffuse radio sources in clusters.

\end{abstract}


%

\section{INTRODUCTION}
\label{intro}

{\it Radio haloes} (also {\it giant radio haloes}, or GRHs) are
diffuse sources found in the central regions of massive
($M_{200}\gtrsim 10^{15}$\mm) galaxy clusters. Extended over volumes
of $\sim$1 Mpc$^3$, these low surface brightness features ($\sim$1
$\mu$Jy arcsec$^{-2}$ at 1.4 GHz) have no obvious optical counterparts
and roughly follow the intracluster medium (ICM) mass
distribution. Their synchrotron steep-spectrum
($S(\nu)\sim \nu^{-\alpha}$; $\alpha>1$) reveals the existence of a
population of relativistic electrons and large-scale magnetic fields
spread throughout the ICM (see, e.g., Feretti et al. \citeyear{fer12}
for a review).

Until recently, GRHs were always discovered in merging clusters
(e.g. van Weeren et al. \citeyear{wer19}). Indeed, in the hierarchical
scenario of cosmic structure formation, it is a common fact that
galaxy groups and subclusters merge together into massive
clusters. These processes release enormous amounts of gravitational
energy (as large as $10^{64}$ erg; Sarazin \citeyear{sar02}) and
induce turbulence in the ICM, which is considered the key mechanism
able to accelerate particles to relativistic energies
(e.g. Brunetti \& Jones \citeyear{bru15}).

However, in the last years this picture got more complicated following
the discovery of diffuse radio emission in several dynamically relaxed
clusters. The first, impacting, case was the extended radio source
found in the cluster CL1821+643 ($z\sim 0.296$; Bonafede et
al. \citeyear{bon14}, Boschin et al. \citeyear{bos18}), whose size,
location and power resemble that of typical GRHs despite the absence
of any merging process responsible for its formation. Indeed, this
case is not unique, since diffuse radio sources not powered by major
mergers have been recently discovered in more clusters in the redshift
range 0.1-0.3. For instance, Abell 2390 and Abell 2261 (Sommer et
al. \citeyear{som17}) are two more examples of cool-core clusters with
Mpc-scale radio sources. Intriguing are also the cases of
PSZ1G139.61+24 and Abell 2142. The first one hosts both a mini-halo (a
feature tipically found in cool-core relaxed systems, e.g. Gitti et
al. \citeyear{git18}) and an underluminous and ultrasteep spectrum
radio halo (Savini et al. \citeyear{sav18a}). Abell 2142, on its hand,
is affected by minor merging activity, as suggested by the cold fronts
detected in its ICM. However, as in the case of PSZ1G139.61+24, it
also exhibits a double-component extended radio emission (Venturi et
al. \citeyear{ven17}), with a spectral steepness increasing in the
outer regions. These two last examples could be hybrid sources, with
mini-halos evolving into GRHs or viceversa (van Weeren et
al. \citeyear{wer19}). It is expected that the number of these
``intermediate'' cases will increase considerably in the near future
thanks to observational facilities like LOFAR (van Haarlem et
al. \citeyear{vha13}).

\begin{figure*}
\centering 
\includegraphics[width=17.5cm]{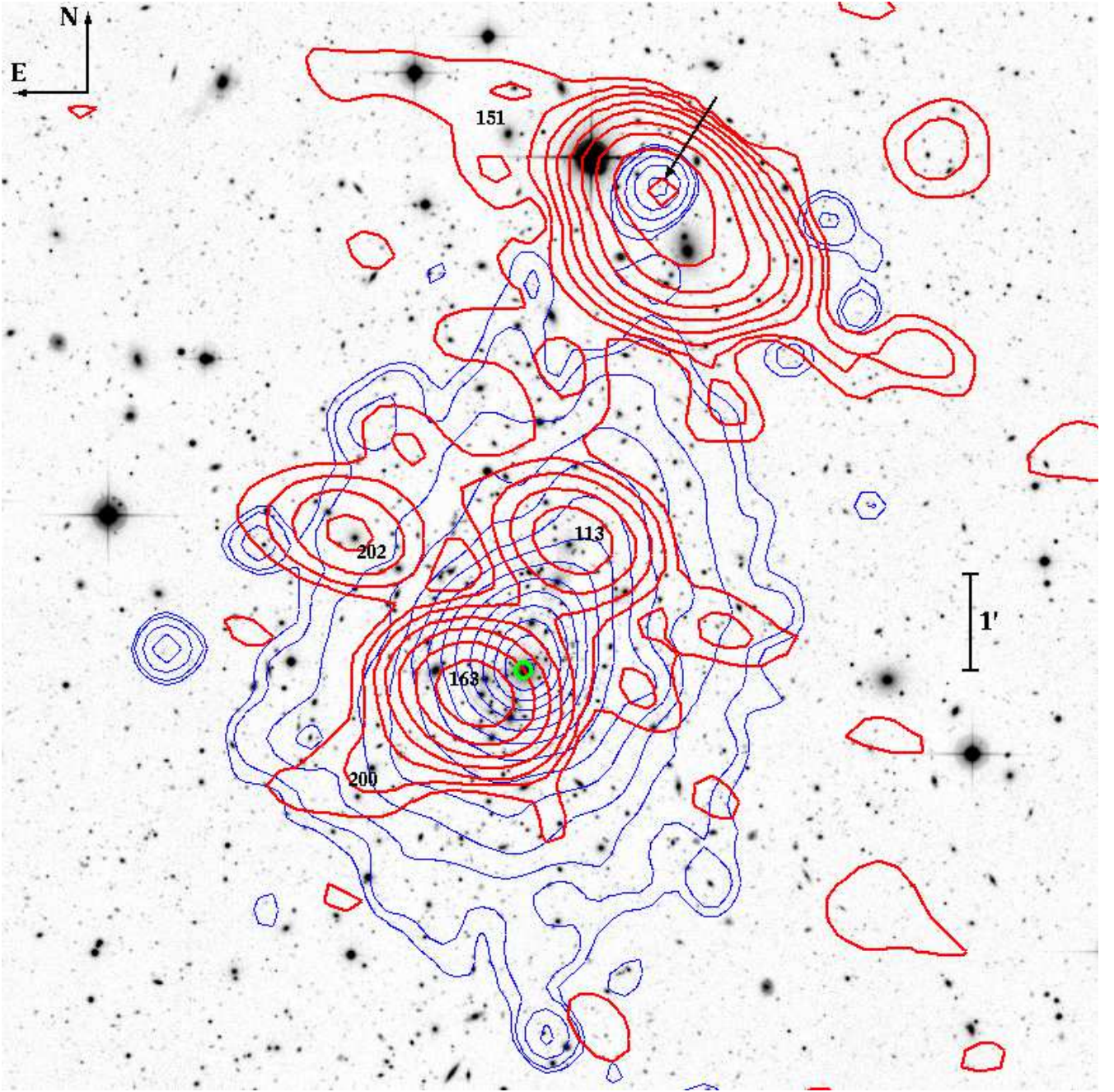}
\caption{
A multiwavelength picture of the cluster A1703. The gray-scale image
in the background corresponds to the optical $r$-band (CFHT/Megaprime
archival data). Blue thin contours show the cluster X-ray emission in
the 0.5-2 keV band (from \textit{Chandra} archival image ID~16126;
Texp: 48 ks). Thick red contours are the contour levels of a VLA 1.4
GHz low-resolution image (courtesy F. Govoni; from archival VLA
observation program AM 469). Numbers highlight member galaxies
mentioned in the text. The green circle is the BCG (ID~141; see
Table~\ref{catalogA1703}).}
\label{figimage}
\end{figure*}

Taking into account this new framework, in this paper we focus on
Abell 1703 (hereafter A1703; Abell \citeyear{abe58}, Abell et
al. \citeyear{abe89}), a system at $z\sim 0.28$ (Allen et
al. \citeyear{all92}) whose dynamical status and radio properties
might look like the ones of CL1821+643. In fact, several authors claim
the possible existence of a radio halo in this cluster. The first hint
comes from Owen et al. (\citeyear{owe99}; see their Table 2), who
include A1703 in a list of nine clusters with diffuse radio
emission. Indeed, A1703 is the brightest source in Owen et
al. sample. More recently, the analysis of archival VLA data at 1.4
GHz showed some evidence of diffuse emission in the central region of
the cluster (Govoni 2018, priv. comm.; see also our
Fig.~\ref{figimage}) despite contamination by radio pointlike sources
(see Rizza et al. \citeyear{riz03} and our
Sect.~\ref{notable}). Finally, Wilber ({\citeyear{wil18}}) highlights
the presence in A1703 of possible radio halo emission at lower
frequencies (120-168 MHz) in the LOFAR Two-Metre Sky Survey
(LoTSS). This evidence is also confirmed by Savini et
al. (\citeyear{sav18b}) in their Figure 8, where the contour levels at
144 MHz from LOFAR suggest the existence of a diffuse source in the
location of the cluster.

\begin{figure*}
\centering 
\includegraphics[width=17.5cm]{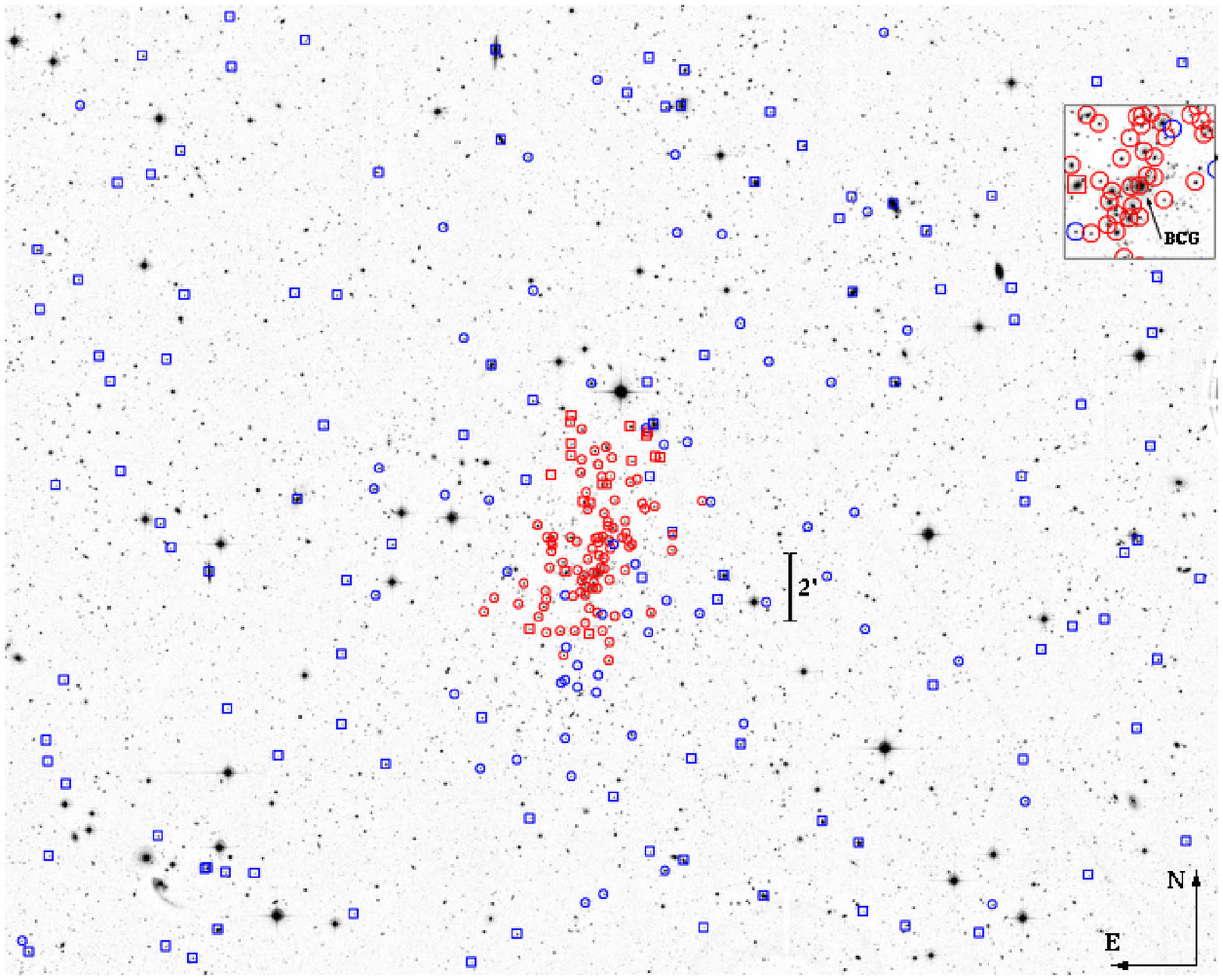}
\caption{
  $r$-band wide field image (CFHT/Megaprime archival data) of the
  cluster A1703 showing the positions of all the galaxies of our
  spectroscopic catalogue. Circles and squares indicate cluster members
  and nonmember galaxies, respectively. Red symbols refer to galaxies
  observed at the TNG (see text and Table~\ref{catalogA1703}). Top
  right inset is a zoom on the central region of the cluster.}
\label{figottico}
\end{figure*}

On the other hand, the picture of A1703 that emerges at other
wavelengths is controversial and suggest conflicting interpretations
about the real dynamical status of this cluster. From the optical
point of view A1703 is one of the richest clusters in the photometric
catalogue of the SDSS (e.g. Koester et al. \citeyear{koe07}) and
presents a dominant giant elliptical cD galaxy in its centre (the
brightest cluster galaxy; hereafter BCG). Moreover, this cluster shows
impressive strong gravitational lensing features (Hennawi et
al. \citeyear{hen08}) which allowed to reconstruct its internal mass
profile. For instance, Limousin et al. ({\citeyear{lim08}), based on
imaging data both from space (HST) and ground (Subaru), identified 13
systems forming highly-magnified images. Thanks to spectroscopic (from
Keck) and photometric redshifts of these images they found that all
the lensing systems can be reproduced by a single NFW (Navarro et
al. \citeyear{nav97}) profile for the dark matter, thus suggesting
that A1703 is a regular, unimodal cluster. A follow-up study by
Richard et al. (\citeyear{ric09}) based on further Keck spectroscopic
measurements for seven multiple sources reinforced the evidence that
A1703 is a relaxed cluster. More studies based on strong and/or weak
gravitational lensing analysis also provided estimates of the cluster
mass and the NFW concentration parameter. Even though the results are
quite discordant, they all coincide that A1703 is a massive system:
$M_{\rm vir}
\simeq 1.1-2\,\times 10^{15}$ \m and $c_{\rm vir}\simeq 3.3-7.1$ 
(Oguri et al. \citeyear{ogu09}, Zitrin et al. \citeyear{zit10}, Oguri
et al. \citeyear{ogu12}).

Such a large mass is also consistent with estimates derived from {\it
Planck} observations of the thermal Sunyaev-Zeldovich effect. In fact,
based on measurements of the Compton parameter $Y$ and the $Y-M_{500}$
scaling relation of Arnaud et al. (\citeyear{arn10}), the Planck
Collaboration (\citeyear{pla16}) reports an hydrostatic mass $M^{\rm
SZ}_{500}=6.76^{+0.35}_{-0.37}\times 10^{14}$\mm, which corresponds to
$M_{\rm vir}\gtrsim 1\times 10^{15}$\m assuming a NFW profile for the
mass distribution.

As for the X-ray band, Piffaretti et al. (\citeyear{pif11}) report a
measurement of the X-ray luminosity of A1703 from \rosat$\,$ data:
$L_{\rm X}(R<R_{500}=1.07$ Mpc)$=5.42\times 10^{44}$ erg s$^{-1}$ in
the 0.1-2.4 keV band. Using eq.~10 of B\"ohringer et
al. (\citeyear{boh14}), this luminosity translates to a mass estimate
$M_{200}\gtrsim 1\times 10^{15}$\m (in our cosmology, see below),
again indicative of a massive cluster. Very recently, Ge et
al. (\citeyear{ge19}) used unpublished \chandra$\,$ archival data to
compute the temperature of the ICM within 0.15-0.75 $R_{500}$:
$kT_{\rm X}=9.63\pm 0.75$ keV. These data show that the ICM is
elongated in the SSE-NNW direction (see contours in
Fig.~\ref{figimage}). However, despite the absence of obvious
bimodality or more complex substructure, Ge et al. (\citeyear{ge19})
also report hints of a disturbed dynamics according to the
measurements of several X-ray morphological parameters (see their
Table~5) and their locations in the morphological planes of Cassano et
al. (\citeyear{cas10}) and Mantz et al. (\citeyear{man15}).

Taking into account the scenarios described by studies based on
gravitational lensing and X-ray data, it is important to definitively
establish whether A1703 is an anomalous relaxed cluster hosting
diffuse radio emission or a common merging system. An exhaustive
analysis of the kinematics of the cluster galaxy population could have
the last word on the true dynamical status of this cluster. Moreover,
this analysis would allow to study the possible presence of a merger
along the line of sight, which would be difficult if not impossible to
detect in the X-ray band. In fact, the spatial and kinematical study
of member galaxies represent an effective tool to reveal substructures
in clusters and put in evidence pre-merging subgroups or merger
remnants (e.g. Boschin et al. \citeyear{bos04}; Boschin et
al. \citeyear{bos13}).

With this context in mind, we used archival spectroscopic data
obtained at the Italian Telescopio Nazionale {\em Galileo} (TNG) in
order to perform the first dynamical analysis of this cluster based on
member galaxies. These data sample the central $\sim$1 Mpc size region
of the cluster characterized by the diffuse X-ray and radio
emissions. More data found in the NED Database, most of which obtained
by Bayliss et al. (\citeyear{bay14}; hereafter B14}) through
spectroscopic measurements with MMT/Hectospec, allowed us to extend
our spectroscopic sample to cover a wider area around the cluster (see
Fig.~\ref{figottico}).

This paper is organized as follows. Sect.~\ref{data} describes the TNG
observations and data reduction and presents the velocity
catalogue. In Sect.~\ref{memb} we describe our member selection
procedure. Sect.~\ref{globalvd} and \ref{clustsub} explain the results
of the analysis of the cluster structure. Finally, in
Sect.~\ref{disc}, we discuss our results and present a portrait of the
dynamical status of A1703.

Unless otherwise stated, we indicate errors at the 68\% confidence
level (hereafter c.l.).  Throughout this paper, we use $H_0=70$ km
s$^{-1}$ Mpc$^{-1}$ in a flat cosmology with $\Omega_{\rm m}=0.3$ and
$\Omega_{\Lambda}=0.7$. In the adopted cosmology, 1\arcm corresponds
to $\sim 253$ kpc at the cluster redshift.

\section{Galaxy data and velocity catalogue}
\label{data}

We used unpublished spectroscopic data stored in the TNG archive
(http://archives.ia2.inaf.it/tng) taken in May 2010 (program
A21TAC\textunderscore 50; PI: F. Gastaldello). These data consist of
five MOS masks mainly sampling the central region of the
cluster. Another mask was taken in June 2016 during a technical
night. In particular, this last mask allowed to obtain the spectrum of
the BCG. For all the six masks we used the LR-B Grism of the
instrument DOLoRes\footnote{http://www.tng.iac.es/instruments/lrs} and
obtained spectra for 131 objects. The total exposure times varied from
3600s to 7200s.

We used standard {\sevensize IRAF} tasks to reduce the spectra and
adopted the cross-correlation technique (Tonry \&
Davis \citeyear{ton79}) to compute redshifts for 104 targets. For 18
galaxies we obtained multiple redshift determinations. They allowed us
to obtain a better estimate for the redshift errors. In particular, we
found that the nominal cross-correlation errors are underestimated and
multiplied them by a factor 2.5 (see Girardi et al. \citeyear{gir11}
for details of the redshift computation and their errors). For another
five galaxies (IDs. 117, 152, 154, 162 and 180; see
Table~\ref{catalogA1703}), their redshifts were computed by measuring
the wavelengths of the emission lines in their spectra.

In order to extend our spectroscopic sample to the outskirts of the
cluster, we searched the NED database for galaxies with known redshift
in the field of A1703. We found 184 objects within a radius of
$\sim$20\arcm from the cluster centre. Most of these objects (177)
come from B14, ten of which are in common with our TNG data. To check
for eventual systematic deviations, we performed a straight-line fit
to the TNG and B14 redshift measurements taking into account the
errors on both data sets (see Chapter 15.3 of Press et
al. \citeyear{pre07}). We find an intercept=$(-2\pm 8)\times 10^{-3}$
and a slope=1.01$\pm$0.03 with a $\chi^2$ probability=0.52. Thus, we
added the remaining 167 B14 galaxies and seven NED galaxies to our
sample.

Our final spectroscopic catalogue includes 278 galaxies. The
field of A1703 is covered by the SDSS. Its galaxy catalogue also
provides us complete photometric information for all the galaxies of
the spectroscopic sample in the magnitude bands $g$, $r$, and $i$.

Table~\ref{catalogA1703} lists the velocity catalogue (see also
Fig.~\ref{figottico}): identification number of each galaxy, ID
(Col.~1); redshift source (Col.~2; T:TNG, L:NED, B:B14); right
ascension and declination, $\alpha$ and $\delta$ (J2000, Col.~3);
(dereddened) SDSS $r$ magnitude (Col.~4); heliocentric radial
velocities, $V=cz_{\sun}$ (Col.~5) with errors, $\Delta V$ (Col.~6).

\input{catalogA1703a1a.tex}

\subsection{Notable galaxies}
\label{notable}

A1703 is dominated by the galaxy ID~141 (the BCG). It is by far the
brightest cluster member, being the magnitude difference with the
second brightest member galaxy $\sim$1.3. Its colours ($g-r=1.58$ and
$r-i=0.59$) well match the red sequences of cluster early-type
galaxies (see our analysis of Sect.~\ref{clust2dphotom}) and,
consistently, its optical spectrum is that of an elliptical galaxy
free of emission lines.

The field of A1703 is populated by several pointlike radio
sources. Rizza et al. (\citeyear{riz03}; see their Table~3) report six
sources from high spatial resolution ($\sim$1.5\arcss) VLA images at
20 cm. The counterparts of five of them are the member galaxies
IDs.~113, 151, 163, 200 and 202 (see Fig.~\ref{figimage}). The sixth
one (n. 5 in Table~3 of Rizza et al. \citeyear{riz03}), at
$\sim$5.7\arcm SW of the BCG, is not listed in our spectroscopic
catalogue. Since its photometric redshift is 0.24$\pm$0.04 (from the
SDSS), it is unclear whether it is a cluster member or a foreground
object. Finally, at $\sim$5\arcm NNW of the BCG there is another
strong radio source whose possible optical counterpart (highlighted by
a black arrow on the top of Fig.~\ref{figimage}) is also visible in
the X-ray {\it Chandra} Image. Due to its faint magnitude ($r\sim$23
from CFHT/Megaprime archival data) we argue that this object is a
background AGN.

\section{Removal of nonmembers}
\label{memb}

\begin{figure}
\centering
\resizebox{\hsize}{!}{\includegraphics{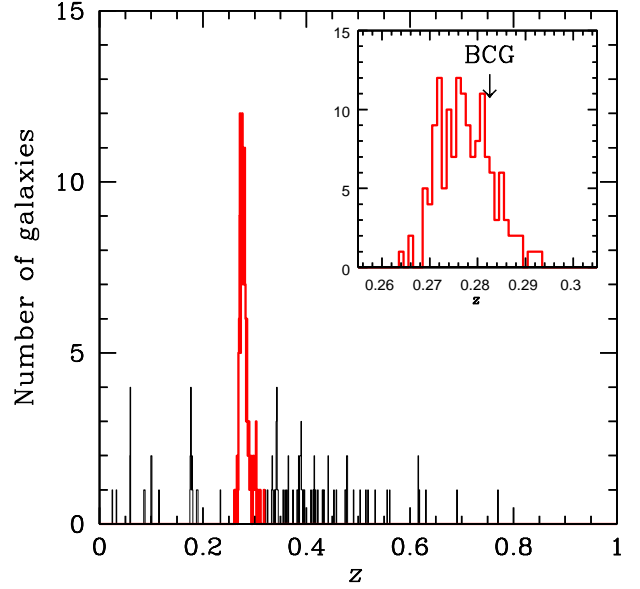}}
\caption
{Redshift histogram of the galaxies of our spectroscopic
  sample. The solid red line histogram highlights the 170 galaxies
  assigned to A1703 by the 1D-DEDICA method. The distribution of the
  final 147 member galaxies with the indication of the BCG redshift is
  shown in the inset plot.}
\label{fighisto}
\end{figure}

\begin{figure}
\centering
\resizebox{\hsize}{!}{\includegraphics{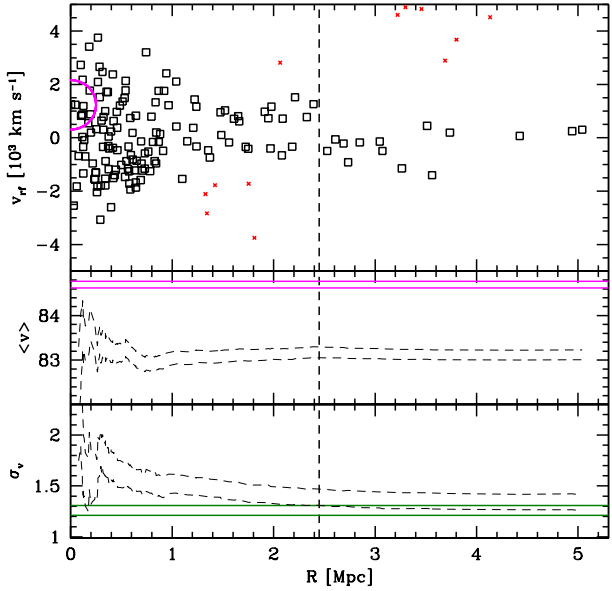}}
\caption
{ {\it Top panel}: Projected phase space diagram for the 147 member
  galaxies (squares) and nonmembers rejected with the shifting gapper
  procedure (small red crosses). Rest-frame LOS velocities are shown
  vs. projected clustercentric distance. The big magenta circle
  highlights the BCG. {\it Middle and Bottom panels}: Integral
  profiles of mean velocity and LOS velocity dispersion, respectively
  (only the one-$\sigma$ error bands are shown). The mean (and
  dispersion) at a given (projected) radius $R$ from the cluster
  centre is estimated by considering all galaxies within that radius
  (the first value computed on the five galaxies closest to the
  centre).  For all the panels, the vertical black dashed line
  indicates $R_{200}$. In the central panel, the horizontal magenta
  lines show the one-$\sigma$ error band of the BCG velocity. In the
  bottom panel, the horizontal green lines show the one-$\sigma$ error
  band of the {\it Chandra} X-ray temperature (from Ge et
  al. \citeyear{ge19}, see Sect.~\ref{intro}) converted to
  $\sigma_{\rm v}$ (see Sect.~\ref{disc} for details).}
\label{figvdprof}
\end{figure}

The removal of nonmember galaxies was performed by using the two-step
method called ``P+G'' (see, e.g. Biviano et al. \citeyear{biv13}),
which combines the 1D adaptive-kernel method DEDICA (1D-DEDICA;
Pisani \citeyear{pis93}) and the ``shifting gapper'' method
(Fadda \citeyear{fad96}). For the centre of A1703 we adopted the
position of the BCG (RA=$13^{\mathrm{h}}15^{\mathrm{m}}05\dotsec24$,
Dec.=$+51\degree 49\arcmm 02.6\arcs$, see
Table~\ref{catalogA1703}). The 1D-DEDICA method detected A1703 as a
peak in the velocity space populated by 170 galaxies (see
Fig.~\ref{fighisto}). Then, we rejected 23 galaxies from this
provisional list of cluster members by using the ``shifting gapper'',
which combines the spatial and velocity information. The final sample
contains 147 member galaxies (87 of which observed at the TNG), whose
projected phase space is shown in Fig.~\ref{figvdprof}.

\section{Global Properties of the velocity distribution}
\label{globalvd}

The biweight routines by Beers et al. (\citeyear{bee90}) provide
robust estimates of the first moments of the velocity distribution.
Our measurement of the mean velocity is
$\left<V\right>=83\,119\pm47$ \ks (or
$\left<z\right>=0.2773\pm0.0002$). The global line-of-sight (LOS)
velocity dispersion is $\sigma_{\rm v}=1324_{-71}^{+88}$ \kss. Based
on the $\sigma_{\rm v}$-$M_{200}$ relation of Munari et
al. (\citeyear{mun13}), inferred from $\Lambda$-cold dark matter
cosmological $N$-body and hydrodynamical simulations, we estimate a
total cluster mass of $M_{200} = (2.2 \pm 0.6)\times 10^{15}$\m within
$R_{200}=2.45 \pm 0.15$ Mpc.

\section{Cluster substructure}
\label{clustsub}

\subsection{1D analysis of the velocity distribution} 
\label{clust1d}

The 1D analysis refers to the study of the higher moments of the
velocity distribution of member galaxies (Fig.~\ref{fighisto}).  In
particular, we find some evidence of deviation from Gaussianity
according to the moments skewness and kurtosis ($\sim 90-95$\% and
$\sim 90-99$\% c.l., respectively). Moreover, there is also marginal
evidence of asymmetry according to the asymmetry index ($\sim 90-95$\%
c.l.; see Bird \& Beers \citeyear{bir93} for details).

Very interestingly, the BCG has a significant ($>99\%$ c.l.) peculiar
velocity $\Delta V_{\rm BCG}$ relative to the cluster mean velocity
according to the Indicator test by Gebhardt \& Beers
(\citeyear{geb91}). In particular, the absolute $\sigma_{\rm
v}$-normalized BCG peculiar velocity is $\left|\Delta V_{\rm
BCG}\right|/\sigma_{\rm v}\sim 1.2$. This value puts the BCG of A1703
in the far tail of the $\left|\Delta V_{\rm BCG}\right|/\sigma_{\rm
v}$ exponential distribution found by Lauer et al. (\citeyear{lau14})
studying a sample of 433 BCGs.

\begin{figure}
\centering
\resizebox{\hsize}{!}{\includegraphics{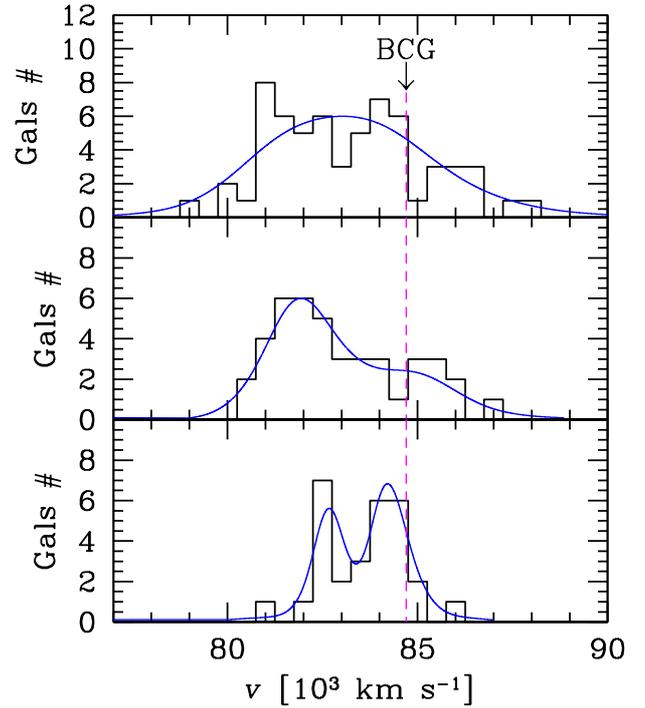}}
\caption
{Velocity histogram of member galaxies in three projected
clustercentric distance intervals: 0-0.5 Mpc ({\it top panel}), 0.5-1
Mpc ({\it middle panel}), and 1-2.5 Mpc ({\it bottom panel}). For all
panels, blue curves show the number-galaxy density in the velocity
space, as provided by the 1D-DEDICA method. The dashed magenta
vertical line highlights the velocity of the BCG.}
\label{fighistogaussmpc}
\end{figure}

We then used the 1D-Kaye Mixture Model method (1D-KMM; Ashman et
al. \citeyear{ash94}). This test quantifies the statistical
significance of bimodality (or more complex structure) in the velocity
distribution with respect to a single Gaussian fit. The results are
negative, i.e. there is no significant evidence of a two- or
three-Gaussian partition. However, if we consider the velocity
histograms obtained selecting member galaxies at different (projected)
clustercentric distances (see Fig.~\ref{fighistogaussmpc}) we unveil a
more complex reality. In fact, while in the centre of the cluster
($R\leq$0.5 Mpc) the velocity distribution is fitted by 1D-DEDICA with
a single peak curve, at larger clustercentric distances signs of
bimodality appear in the form of a fitted asymmetric curve. Then, for
$R\gtrsim 1$ Mpc ($\sim 0.4\,R_{200}$), the velocity distribution is
clearly described by a two-peak curve. The significance of the two
peaks, at $\sim$82700 and $\sim$84200 \kss, is $>$99.4\%
c.l. according to 1D-DEDICA.

\subsection{2D analysis of the galaxy distribution}

\begin{figure}
\centering
\resizebox{\hsize}{!}{\includegraphics{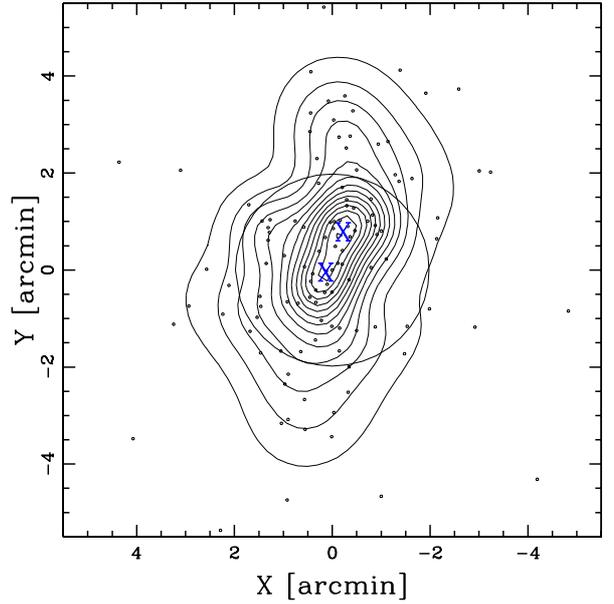}}
\caption
{ Spatial distribution of the spectroscopic cluster members in the
  central region of A1703 with, superimposed, the isodensity contour
  levels obtained with the 2D-DEDICA method. The blue crosses indicate
  the locations of the two galaxy peaks 'S' and 'N' (see also
  Table~\ref{tabdedica2dz}). The plot is centred on the cluster centre
  (the BCG) and circle contains the cluster within a radius equal to
  0.5 Mpc.}
\label{figk2z}
\end{figure}

\begin{figure}
\centering
\resizebox{\hsize}{!}{\includegraphics{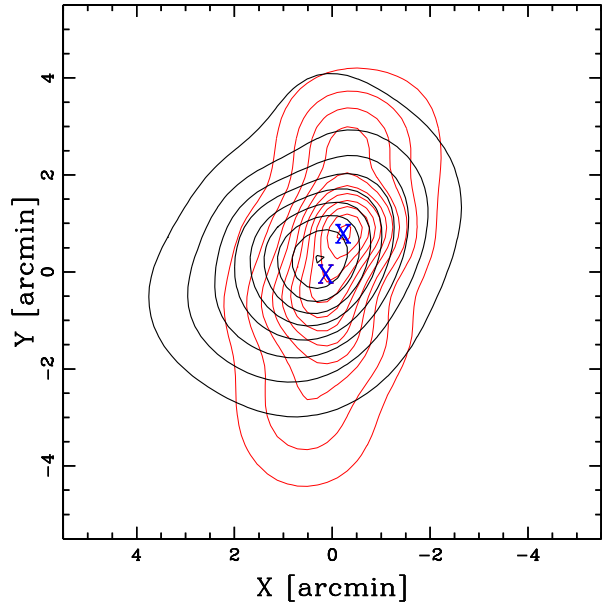}}
\caption
{Isodensity contour map of the spectroscopic cluster members
  according to the 2D-DEDICA method. Black thick and red thin contours
  refer to bright ($r\leq 20$) and faint members, respectively (see
  text). The blue crosses indicate the locations of the galaxy peaks
  'S' and 'N' (see Table~\ref{tabdedica2dz}). The plot is centred on
  the cluster centre (the BCG).}
\label{figk2zm20}
\end{figure}

About the analysis of the 2D spatial distribution of the spectroscopic
member galaxies, we employed the 2D adaptive-kernel method of Pisani
et al. (\citeyear{pis96}, hereafter 2D-DEDICA). The results are shown
in Fig.~\ref{figk2z}. The cluster is elongated along the SSE-NNW
direction, in a similar way to the X-ray isophotes (see
Fig.~\ref{figimage}). Moreover, this test detects two dense galaxy
peaks separated by only $\lesssim$1\arcmm. The peak 'S' (see
Table~\ref{tabdedica2dz}) is located close to the BCG (at $\sim$8\arcs
ESE), the peak 'N' is found at $\sim$9\arcs SE of the galaxy
ID~120. The eventual presence of luminosity segregation in a
galaxy cluster can be a sign of disturbed dynamics (e.g. Maurogordato
et al. \citeyear{mau11}; see the discussion in Sect.~\ref{disc}). This
motivates the exploration of the 2D galaxy distribution in different
magnitude ranges. Very interestingly, running 2D-DEDICA only on the
bright spectroscopic members ($r\leq 20$) the result is completely
different with respect to the analysis of the whole spectroscopic
sample: the two peaks and the elongation disappear. The distribution
of bright members is now much more circular (see Fig.~\ref{figk2zm20})
with a centre close to the BCG and to the peak 'S'. Moreover, the
distribution of faint spectroscopic members shows a peak in
correspondence of the peak 'N'.

\input{tabdedica2dz.tex}

\subsubsection{Analysis of the photometric sample}
\label{clust2dphotom}

We are aware that the spectroscopic sample suffers from magnitude
incompleteness. This is caused by constraints in the production
process of the TNG MOS masks and the positioning of Hectospec fibers
for the galaxies observed by B14.

\begin{figure}
\centering 
\resizebox{\hsize}{!}{\includegraphics{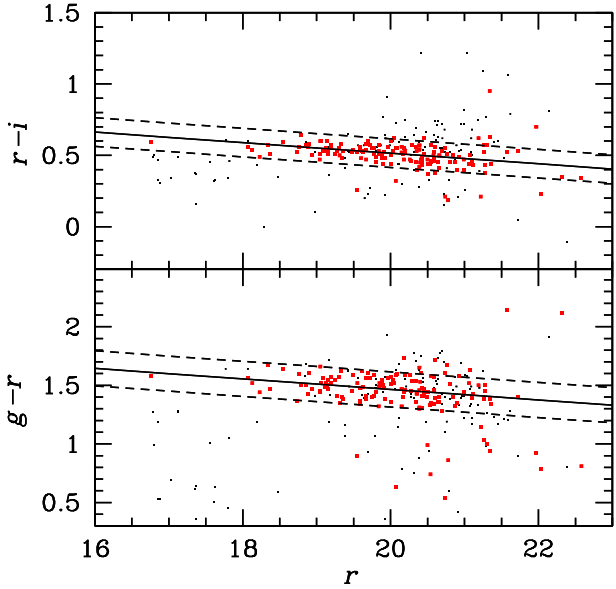}}
\caption
    {{\it Upper panel}: $r-i$ vs. $r$ diagram for galaxies with
available spectroscopic data. Red squares are cluster members, while
black points represent field galaxies. The solid line shows the
best-fit CMR as computed from cluster members; the dashed lines are
drawn at $r-i\pm$0.1 mag from the CMR (see text). {\it Lower panel}:
$g-r$ vs. $r$ diagram for galaxies with available spectroscopic
data. As above, the solid line shows the best-fit CMR and the dashed
lines are drawn at $g-r\pm$0.15 mag from the CMR.}
\label{figcm}
\end{figure}

The SDSS photometry of the cluster field is deep enough to help us
alleviate our incompleteness problems. In particular, we select likely
members on the basis of both ($r-i$ vs. $r$) and ($g-r$ vs. $r$)
colour-magnitude relations (hereafter CMRs). The CMRs allow us to
identify the cluster ``red'' early-type galaxies (i.e., the dominant
cluster population, Dressler \citeyear{dre80}) and to reduce the
contamination by nonmember galaxies. We determine the CMRs by applying
the 2$\sigma$-clipping fitting procedure to the cluster members and
obtain $r$--$i$=1.255-0.037$\times r$ and $g$--$r$=2.365-0.045$\times
r$ (see Fig.~\ref{figcm}). Then, within the photometric catalogue we
consider as likely ``red'' cluster members the galaxies with colour
indexes $r$--$i$ and $g$--$r$ within 0.1 mag and 0.15 mag (i.e. the
1$\sigma$-error associated to the fitted intercept) of the respective
CMRs.

\begin{figure}
\centering
\resizebox{\hsize}{!}{\includegraphics{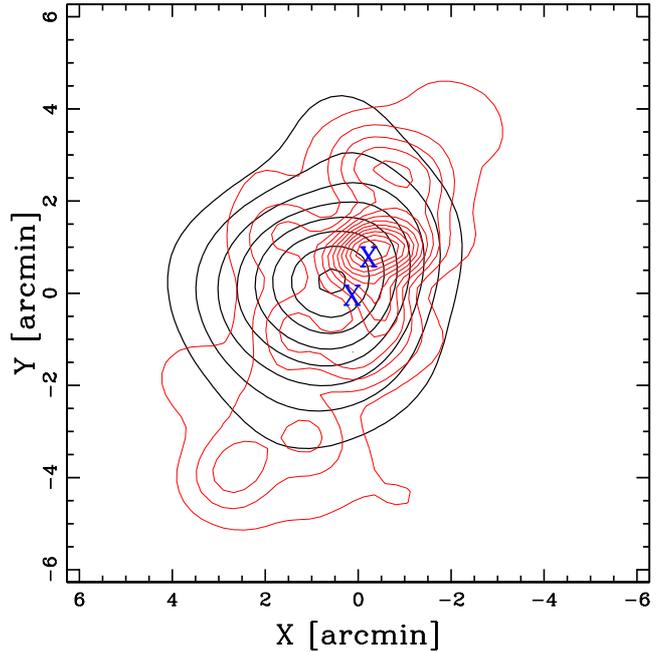}}
\caption{
  Isodensity contour map of the photometric likely cluster members
  according to the 2D-DEDICA method. Black thick and red thin contours
  refer to bright ($r\leq 20$) and faint members, respectively (see
  text). The blue crosses indicate the locations of the galaxy peaks
  'S' and 'N' (see Table~\ref{tabdedica2dz}). The plot is centred on
  the cluster centre (the BCG).}
\label{figk2rigrm20}
\end{figure}

Fig.~\ref{figk2rigrm20} shows the contour map of the likely cluster
members according to 2D-DEDICA. Again, we find that the distribution
of bright galaxies ($r\leq 20$) shows just one peak and is only mildly
elongated. Only if we consider fainter members (galaxies with
$20<r\leq 21.5$) we recover the SSE-NNW elongation with a maximum
galaxy density close to the 'N' peak of Table~\ref{tabdedica2dz}.

This luminosity segregation is not a unique feature of A1703. In
Sect.~\ref{disc} we discuss this result in more detail.

\subsection{3D analysis: combining velocity and position information}
\label{clust3d}

As for the 3D analysis, we employed different tools to search for a
correlation between velocity and position information, which would be
a clear sign of real substructures in the cluster.

First, we searched for an eventual velocity gradient by performing a
multiple linear regression fit to the cluster velocity field (den
Hartog \& Katgert \citeyear{den96}). We find marginal evidence (at the
$\sim$92\% c.l.) of a velocity gradient with PA=$180\pm29$ degrees in
the sample of the 147 spectroscopic cluster members. In particular,
the southern region is populated by higher velocity galaxies.

Then, over the same sample we apply the classical $\Delta$-test
(Dressler \& Schectman \citeyear{dre88}, hereafter DS-test), which
quantifies substructure searching for subsystems whose mean velocities
and/or dispersions deviate from the global cluster values.  Very
significant substructure (at $>99.9$\% c.l., checked by running a
Monte Carlo shuffling of the galaxy velocities; Dressler \&
Schectman \citeyear{dre88}) is found in A1703 both with the
``canonical'' DS-test and its modified version, which consider only
the local mean velocity as kinematical indicator (see also, e.g.,
Girardi et al. \citeyear{gir10} for more details). Again, high
velocity galaxies tend to populate the southern region of the cluster.

\begin{figure}
\centering 
\resizebox{\hsize}{!}{\includegraphics{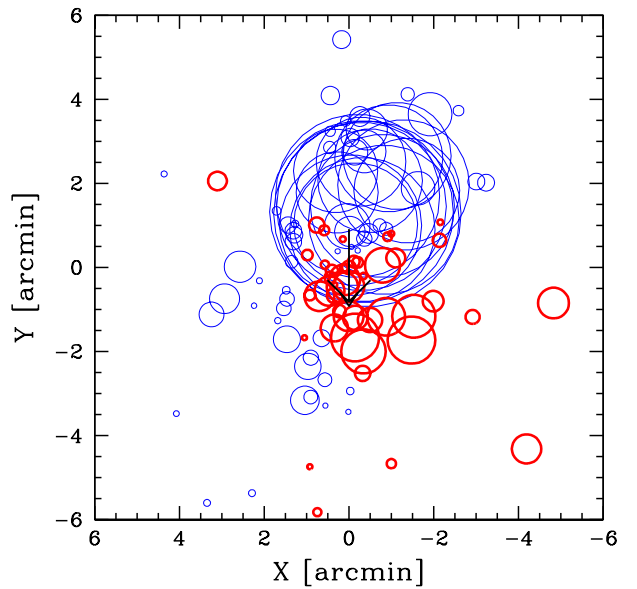}}
\caption
{Spatial distribution of the cluster members in a
  12\arcmm$\times$12\arcm box, each marked by a circle: the larger the
  circle, the larger is the deviation of the local mean velocity from
  the global mean velocity. In particular, thin/blue and thick/red
  circles show galaxies whose local value is smaller/larger than the
  global one according to the modified DS-test. The plot is centred on
  the cluster centre (the BCG). The black big arrow shows the
  direction of the velocity gradient (see text).}
\label{figdssegno6v}
\end{figure}

Later, we resorted to the ``hierarchical tree'' (Htree) algorithm
developed by Serna \& Gerbal (\citeyear{ser96}; see also Adami et
al. \citeyear{ada18} and Girardi et al. \citeyear{gir19} for recent
applications). We apply it to the catalogue of 147 member
galaxies. The method computes the relative binding energies of cluster
galaxies and performs a hierarchical clustering analysis to detect
galaxy subsystems.

\begin{figure*}
\centering 
\includegraphics[width=10cm,angle=90]{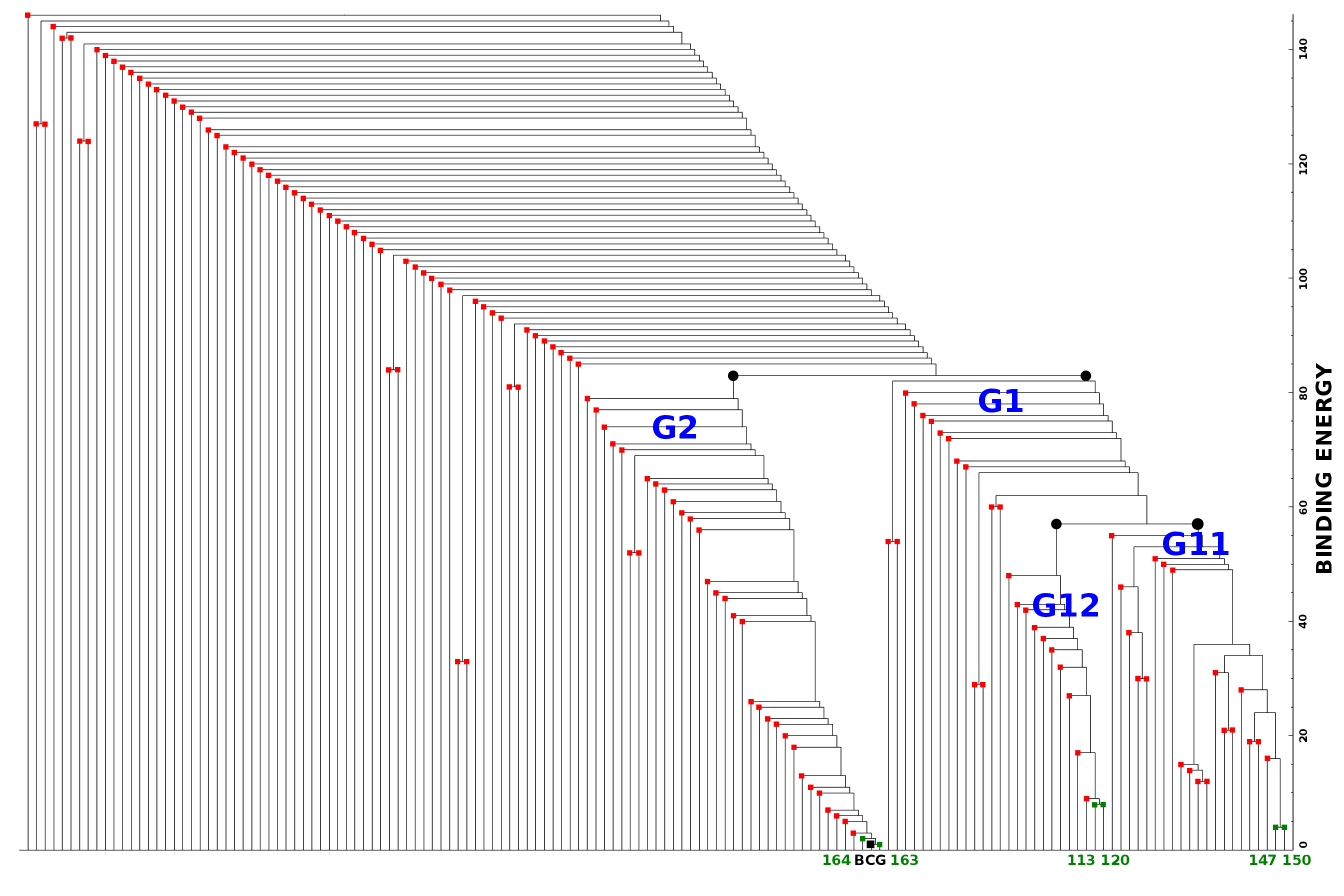}
\caption{
      Dendrogram obtained through the Serna \& Gerbal algorithm
      adopting $M/L_{r}$=150 \mmlr. The y-axis is the binding
      energy, here in arbitrary units with the lowest energy levels on
      the bottom. Labels indicate prominent galaxies and structures.}
\label{figHT}
\end{figure*}

The results of the Htree test are convincing: the cluster hosts two
main substructures. In the centre of the dendrogram of
Fig.~\ref{figHT} we find the group G2. It contains 35 galaxies and, in
particular, the BCG. On the right the group G1 is the most prominent
structure.  It contains 47 galaxies and is itself substructured in two
groups, G12 (12 galaxies) and G11 (21 galaxies). We run the Htree
test by assuming a constant value of $M/L_{r}$=150 \mmlr for the
mass-to-light ratio of galaxies, but the results are quite robust
against the adopted value of $M/L_{r}$.

\input{tabhtree.tex}

In Table~\ref{tabhtree} we report the properties of the subclumps
found with the Htree method.

\begin{figure}
\centering 
\resizebox{\hsize}{!}{\includegraphics{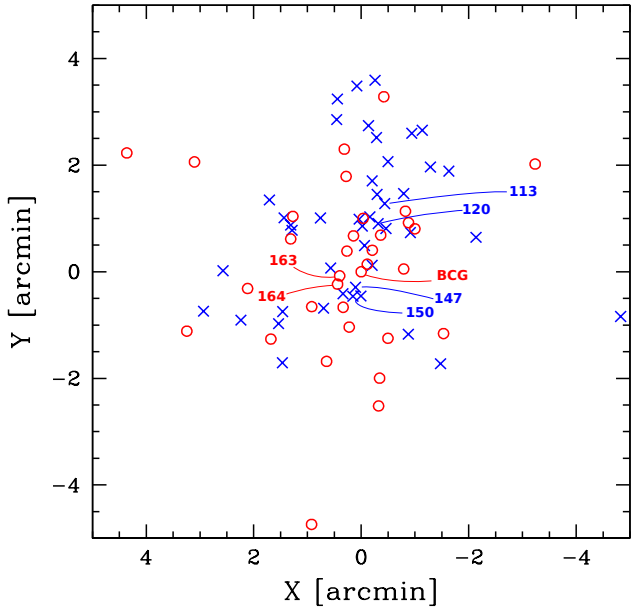}}
\caption{Member galaxies belonging to the groups G1 (blue crosses) and G2 (red circles) in a 10\arcmm$\times$10\arcm box centred on the BCG. Labels highlight galaxies at the lowest binding energy levels of their respective groups (see Fig.~\ref{figHT}).}
\label{figxyht}
\end{figure}

In Fig.~\ref{figxyht} we show the positions of the galaxies belonging
to the groups G1 and G2. G1 galaxies have lower velocities and
populate mainly the northern region of the cluster.

Note that the velocity dispersions of the groups of
Table~\ref{tabhtree} are probably underestimated, since they do not
include all the galaxies of the cluster and could just be the cores of
more massive structures. To overcome this point we used the 3D version
of the KMM test (3D-KMM). In particular, we used the galaxy
assignments of the groups G1 and G2 as a first guess when fitting two
groups. The algorithm fits a two groups partition at the 98.4\%
c.l. The results for the two groups are reported in
Table~\ref{tabkmm3d}.  Based on the estimates of the velocity
dispersions of the two groups, we find from Munari et
al. (\citeyear{mun13}): $R_{200}=1.45\pm 0.1$ Mpc and $M_{200}=(4.6\pm
1.6)\times 10^{14}$\m for KMM3D1, $R_{200}=1.6\pm 0.2$ Mpc and
$M_{200}=(5.8\pm 2.6)\times 10^{14}$\m for KMM3D2. Therefore, the
total mass would be $M_{200}= (10.4\pm 3)\times 10^{14}$\mm.

\input{tabkmm3d.tex}

Finally, the substructure found by the Htree test is confirmed by the
analysis performed with the three dimensional adaptive-kernel method
of Pisani (\citeyear{pis93}, \citeyear{pis96}; 3D-DEDICA). The results
are reported in Table~\ref{tabdedica3dz}, where the three groups
DED3D1, DED3D2 and DED3D3 correspond to the groups G2, G12, and G11
detected by the Htree method.

\input{tabdedica3dz.tex}

\begin{figure}
\centering 
\resizebox{\hsize}{!}{\includegraphics{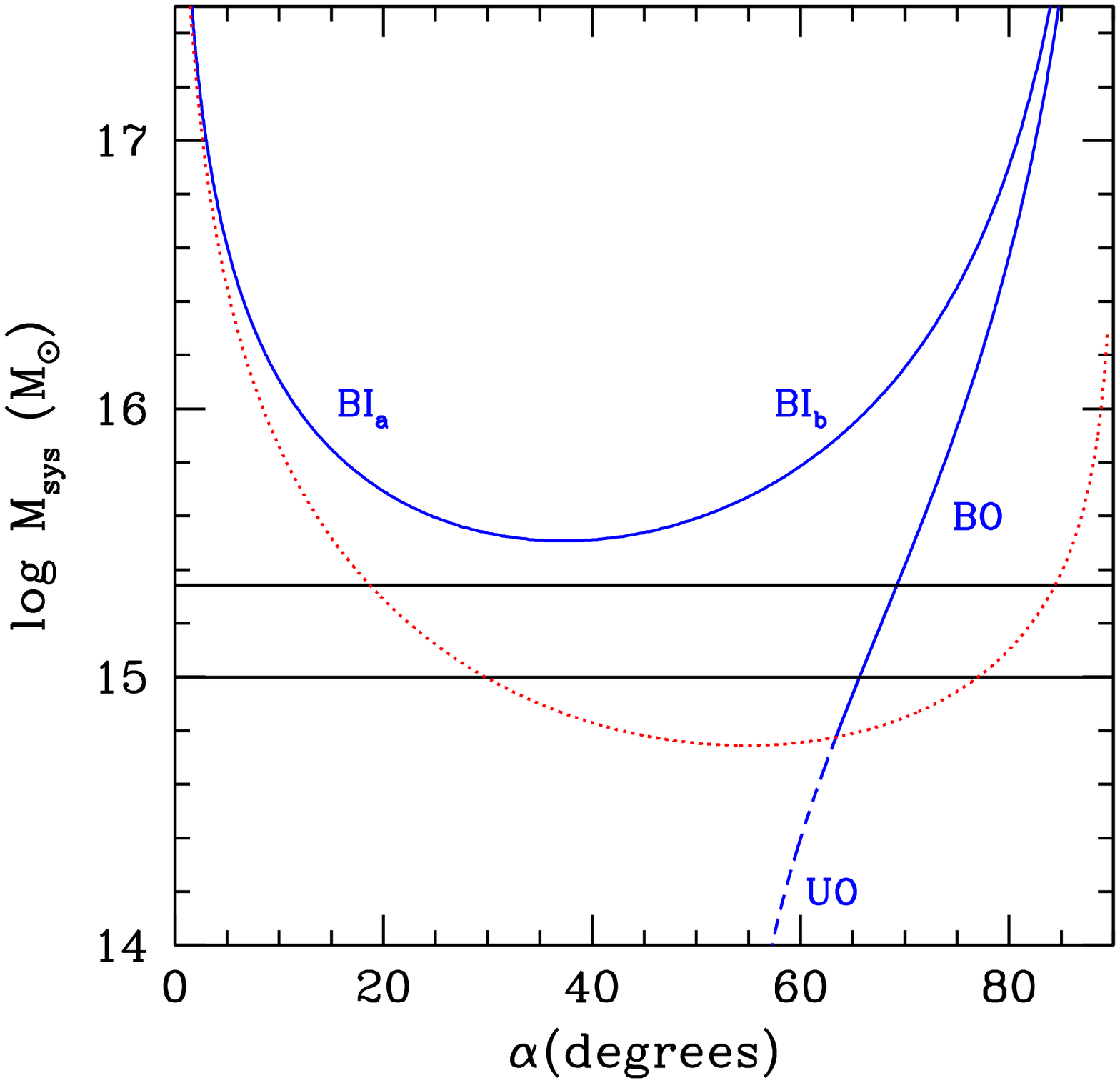}}
\caption{
Bound and unbound solutions of the two-body model applied to the
subclusters G1 and G2 (thick solid and dashed curves,
respectively). The cluster mass is plotted versus the projection
angle. Acceptable solutions are the curves intersecting the rectangle
defined by the observational values of the cluster mass (bordered by
the two horizontal lines). Regions above and below the thin dashed
curve are the loci of bound and unbound solutions, respectively,
according to the Newtonian criterion of gravitational binding
(e.g. Beers et al. \citeyear{bee82}).}
\label{figbim}
\end{figure}

\section{Discussion and conclusions}
\label{disc}

The value of the global velocity dispersion $\sigma_{\rm
v}=1324_{-71}^{+88}$ \ks (in agreement with the estimate of B14) is
typical of a massive cluster and is consistent with the X-ray
temperature $kT_{\rm X}=9.63\pm 0.75$ keV measured by Ge et
al. (\citeyear{ge19}) under the assumption of energy density
equipartition between galaxies and ICM. In fact, we find $\beta_{\rm
spec}=\sigma_{\rm v}^2/(kT_{\rm X}/\mu m_{\rm
p})=1.10_{-0.15}^{+0.17}$.

A value of $\beta_{\rm spec}\sim 1$ is not anomalous for a dynamically
relaxed cluster as A1703 was thought to be until very
recently. However, we find convincing evidence that this cluster is
experiencing a merger of two or more subclumps. The first hint of a
disturbed dynamics comes from the 1D analysis of the galaxy velocity
distribution. Even if its statistical moments suggest only marginal
evidence of deviation from the Gaussianity, the velocity distributions
of member galaxies at various clustercentric distances (see
Sect.~\ref{clust1d} and Fig.~\ref{fighistogaussmpc}) point out the
possible existence of two galaxy populations with different mean
velocities (separated by $\sim 2000$ \kss) but well mixed in the
central ($R\lesssim 1$ Mpc) region of the cluster. However, the most
compelling argument in favour of an ongoing merger in A1703 comes from
the very significant peculiar velocity of the BCG (e.g. Martel et
al. \citeyear{mar14}). This is in sharp contrast with what is usually
found in regular clusters, where the dominant galaxy is well placed at
the peak of the global velocity distribution (as in the case, e.g., of
CL1821+643; Boschin et al. \citeyear{bos18}).

About the 2D analysis of the galaxy distribution, the 2D-DEDICA method
applied on the whole set of 147 member galaxies shows that the cluster
is elongated in the SSE-NNW direction (Fig.~\ref{figk2z}) with two
closely spaced peaks (see Table~\ref{tabdedica2dz}), another sign that
at least two subclumps are in the process of merging. Indeed, if we
consider only bright members with $r\leq 20$, their spatial
distribution is much more circular and exhibits only one peak close to
the BCG and the peak 'S'. We remark that this luminosity segregation
is confirmed also by the 2D analysis of the photometric members
distribution, thus it does not seem an artifact of the incompleteness
of the spectroscopic sample. Moreover, note that the peak 'N' and the
density peak observed in the distribution of faint member galaxies
(Figs.~\ref{figk2zm20} and \ref{figk2rigrm20}) coincide with a
secondary mass peak detected by Zitrin et al. (\citeyear{zit10}, see
their Fig.~7) from their lensing analysis. Such a spatial segregation
between more and less luminous galaxies was found, for instance, in
the Coma cluster (Biviano et al. \citeyear{biv96}) and, in particular,
in the merging cluster Abell 209 (Mercurio et al. \citeyear{mer03a}),
where bright photometric members are located around the dominant
galaxy, while fainter galaxies seem to trace the whole cluster
structure in agreement with the X-ray cluster morphology. As in the
case of Abell 209 (see also Mercurio et al. \citeyear{mer03b}), the
segregation observed in A1703 could be interpreted as the result of a
merging process caught after the core-core passage, where luminous
galaxies trace the remnants of a pre-merging clump hosting the BCG.

Finally, the combined analysis of the velocity and spatial information
provides the ultimate evidence of subclustering in A1703. In
particular, the classical DS-test detects very significant
substructure, with high velocity galaxies mostly located south of the
BCG (Fig.~\ref{figdssegno6v}). The Htree-test makes it possible to
identify two main subclusters: G1 and G2 (Fig.~\ref{figxyht} and
Table~\ref{tabhtree}). G2 hosts the BCG and is populated by high
velocity galaxies. G1 exhibits a lower mean velocity and is itself
substructured in two subgroups: G11 and G12. G12, in particular, hosts
in the bottom of its potential well the galaxy ID~120, placed very
close to the 'N' peak (Table~\ref{tabdedica2dz}). Comfortingly, the
existence of G11, G12 and G2 is independently confirmed also by the 3D
version of the adaptive kernel technique (see
Table~\ref{tabdedica3dz}). The spatial distribution of G1 and G2
galaxies (see Fig.~\ref{figxyht}) explains the (marginal) evidence of
a velocity gradient (Sect.~\ref{clust3d}) in the direction North-South
and is in agreement with the results of the DS-test.

Using the galaxies assigned to G1 and G2 as a first guess when fitting
a two groups partition, the 3D-KMM test divides the spectroscopic
sample in two subclumps (Table~\ref{tabkmm3d}). Based on their
velocity dispersions we estimate a total mass $M_{200}\sim 1\times
10^{15}$\m which should be interpreted as a lower limit for the mass
of the cluster, being the upper limit $M_{200}\sim 2\times 10^{15}$\m
derived from the measurement of the global velocity dispersion
(Sect.~\ref{globalvd}).

Now, we can adopt the above estimates for the mass of A1703 to study
in more detail the merger of the two main subclusters G1 and G2. In
particular, we use an analytical two-body model (see, e.g., Lubin et
al. \citeyear{lub98} for the details of the method) which is based on
the following parameters: $M_{\rm sys}$, the total mass of the system,
$D$, the projected distance between G1 and G2, and $\Delta V_{\rm
rf}$, the relative LOS velocity (in the rest frame). From the biweight
estimate of the centres of G1 and G2 we assume $D=0.46$ Mpc, while
their relative velocity is $\Delta V_{\rm rf}\sim 2000$ \kss. If we
also assume that we are seeing the two subclusters at $t=0.3$ Gyr
after the core crossing, which is a typical time suggested by the
presence of the suspected radio halo in A1703 (e.g., Brunetti et
al. \citeyear{bru09}), we can plot a $M_{\rm sys}$ versus $\alpha$
(the projection angle between the plane of the sky and the vector
defined by the centres of G1 and G2) graph for three types of
solutions: Bound outgoing (expanding) solutions (BO), bound incoming
(collapsing) solutions (BI), and unbound outgoing solutions (UO). The
results are reported in Fig.~\ref{figbim}. The only acceptable
solutions in the estimated range of mass of A1703 are BO solutions
with $\alpha\sim 65-70$ degrees, i.e. we are seeing a merger quite
close to the LOS. This is consistent with the hypothesis that the
possible extended radio emission observed in the centre of the cluster
is a radio halo. In fact, an eventual central radio relic produced by
a shock wave propagating along the LOS is a rare event, being a relic
preferentially observed in mergers occurring near the plane of the sky
and in the outermost regions of a cluster (e.g. Vazza et
al. \citeyear{vaz12}, Golovich et al. \citeyear{gol19}).

In conclusion, the observational scenario suggests that A1703 is a
massive cluster undergoing strong dynamical evolution, with two or
three subclusters involved in a merging process. This evidence arises
despite the cluster appears only slightly elongated in the X-rays, but
this is not unusual. Indeed, a non-relaxed state is observed in
$\sim$70\% of clusters at $z\sim 0.2$ even in the presence in some
cases of a regular shape in the X-ray imaging data (Smith et
al. \citeyear{smi05}). Thus, our results are in agreement with the
disturbed dynamics found by Ge et al. (\citeyear{ge19}) through the
study of several X-ray morphological indicators.

In this merging context, the possible presence of a radio halo in the
central regions of the cluster would not be surprising, thus we rule
out the hypothesis that A1703 could constitute a new intriguing case
of a rare relaxed cluster hosting diffuse radio emission.

Finally, our results are in marked contrast with respect to the
picture of an unimodal, dynamically relaxed cluster painted by
previous studies based on gravitational lensing (see
Sect.~\ref{intro}). Since cluster mass profiles inferred from
gravitational lensing suffer from complicated lens geometry (see
Narayan \& Bartelmann \citeyear{nar96} for a general review), if A1703
were a rare relaxed cluster at intermediate redshift it would be of
great value to test structure formation in the $\Lambda$CDM paradigm
through the study of its regular mass distribution. On the contrary,
the subclustering we find in A1703 seems to rule out this possibility.

\section*{Acknowledgements}

We thank the anonymous referee for his/her stimulating comments
and suggestions.

We are in debt with Federica Govoni for the VLA radio image she kindly
provided us. We also thank Luigina Feretti for useful suggestions
and discussions.

M.G. acknowledges financial support from the grant MIUR PRIN 2015
``Cosmology and Fundamental Physics: illuminating the Dark Universe
with Euclid'' and from the University of Trieste through the program
``Finanziamento di Ateneo per progetti di ricerca scientifica - FRA
2018''.

This publication is based on observations made on the island of La
Palma with the Italian Telescopio Nazionale {\it Galileo}, which is
operated by the Fundaci\'on Galileo Galilei-INAF (Istituto Nazionale
di Astrofisica) and is located in the Spanish Observatorio of the
Roque de Los Muchachos of the Instituto de Astrof\'isica de Canarias.

This research has also benefited from the galaxy catalogue of the
Sloan Digital Sky Survey (SDSS). The SDSS web site is
http://www.sdss.org/, where the list of the funding organizations and
collaborating institutions can be found.

This research has made use of the NASA/IPAC Extragalactic
Database (NED), which is operated by the Jet Propulsion Laboratory,
California Institute of Technology, under contract with the National
Aeronautics and Space Administration.

This research has also made use of {\sevensize IRAF}. This
package (ASCL code record 9911.002) is distributed by the National
Optical Astronomy Observatory, which is operated by the Association of
Universities for Research in Astronomy (AURA) under a cooperative
agreement with the National Science Foundation.

\input{catalogA1703a1b.tex}

\bsp	
\label{lastpage}
\end{document}

%% file: catalogA1703a1a.tex

\begin{table}
        \caption[]{Velocity catalogue of 278 spectroscopically measured
          galaxies in the field of A1703. IDs in italics refer to nonmember galaxies. Galaxy ID~141 (in boldface) is the BCG.}
         \label{catalogA1703}
              $$ 
           \begin{array}{r c c c r r}
            \hline
            \noalign{\smallskip}
            \hline
            \noalign{\smallskip}

\mathrm{ID} & \mathrm{Source} & \mathrm{\alpha},\mathrm{\delta}\,(\mathrm{J}2000)  & r& V\,\,\,\,\,&\mathrm{\Delta}V\\
  & &                 & &\mathrm{(\,km}&\mathrm{s^{-1}\,)}\\
            \hline
            \noalign{\smallskip}  

\textit{001} & \mathrm{B}           &         13\ 13\ 13.66,+51\ 48\ 48.1      & 21.09& 142509&   69 \\
\textit{002} & \mathrm{B}           &         13\ 13\ 16.19,+52\ 03\ 34.4      & 21.46& 166775&   96 \\
\textit{003} & \mathrm{B}           &         13\ 13\ 16.50,+51\ 41\ 17.8      & 20.03& 116430&   27 \\
\textit{004} & \mathrm{B}           &         13\ 13\ 21.13,+51\ 57\ 26.5      & 20.64& 126545&   36 \\
\textit{005} & \mathrm{B}           &         13\ 13\ 21.59,+51\ 46\ 30.6      & 18.98&  70250&   30 \\
\textit{006} & \mathrm{B}           &         13\ 13\ 22.20,+51\ 55\ 51.6      & 20.70&  90154&   54 \\
\textit{007} & \mathrm{B}           &         13\ 13\ 22.63,+51\ 52\ 35.9      & 20.70& 107578&   75 \\
\textit{008} & \mathrm{B}           &         13\ 13\ 25.19,+51\ 49\ 53.9      & 18.84& 115294&   18 \\
\textit{009} & \mathrm{B}           &         13\ 13\ 25.56,+51\ 44\ 31.7      & 20.73& 114710&   39 \\
\textit{010} & \mathrm{B}           &         13\ 13\ 27.59,+51\ 49\ 33.9      & 20.61& 115480&  111 \\
\textit{011} & \mathrm{B}           &         13\ 13\ 31.36,+51\ 47\ 39.4      & 20.99&  86817&   66 \\
\textit{012} & \mathrm{B}           &         13\ 13\ 32.20,+52\ 03\ 01.7      & 20.67& 147285&   39 \\
\textit{013} & \mathrm{B}           &         13\ 13\ 34.59,+51\ 40\ 21.1      & 21.18& 155562&   54 \\
\textit{014} & \mathrm{B}           &         13\ 13\ 35.50,+51\ 53\ 48.3      & 21.72& 116685&   33 \\
\textit{015} & \mathrm{B}           &         13\ 13\ 37.34,+51\ 47\ 28.3      & 21.10& 189304&   57 \\
\textit{016} & \mathrm{B}           &         13\ 13\ 43.03,+51\ 46\ 48.6      & 20.64& 124402&   54 \\
\textit{017} & \mathrm{B}           &         13\ 13\ 45.96,+51\ 51\ 02.8      & 20.27& 114620&   63 \\
         018 & \mathrm{B}           &         13\ 13\ 46.21,+51\ 42\ 28.5      & 19.13&  83693&   33 \\
\textit{019} & \mathrm{B}           &         13\ 13\ 46.53,+51\ 43\ 39.1      & 20.76& 125199&   33 \\
\textit{020} & \mathrm{B}           &         13\ 13\ 46.65,+51\ 51\ 46.1      & 19.90& 154438&   27 \\
\textit{021} & \mathrm{B}           &         13\ 13\ 47.76,+51\ 56\ 13.8      & 20.12& 101459&   51 \\
\textit{022} & \mathrm{B}           &         13\ 13\ 48.24,+51\ 57\ 09.2      & 20.97& 116841&   84 \\
\textit{023} & \mathrm{B}           &         13\ 13\ 51.79,+51\ 59\ 46.9      & 19.49& 102676&   33 \\
         024 & \mathrm{B}           &         13\ 13\ 52.32,+51\ 39\ 31.1      & 19.85&  83369&   39 \\
\textit{025} & \mathrm{B}           &         13\ 13\ 54.87,+51\ 38\ 42.0      & 20.46&  87818&   45 \\
         026 & \mathrm{B}           &         13\ 13\ 58.50,+51\ 46\ 30.0      & 19.59&  82845&   30 \\
\textit{027} & \mathrm{B}           &         13\ 14\ 01.35,+51\ 57\ 06.0      & 23.47&  88999&   84 \\
\textit{028} & \mathrm{B}           &         13\ 14\ 03.13,+51\ 45\ 47.5      & 20.70& 143175&   69 \\
\textit{029} & \mathrm{B}           &         13\ 14\ 04.11,+51\ 58\ 47.1      & 17.61&  17850&  150 \\        
         030 & \mathrm{B}           &         13\ 14\ 07.66,+51\ 55\ 57.6      & 20.65&  82893&   60 \\
\textit{031} & \mathrm{B}           &         13\ 14\ 08.32,+51\ 38\ 55.1      & 20.28& 143490&   42 \\
\textit{032} & \mathrm{B}           &         13\ 14\ 09.91,+51\ 54\ 27.6      & 17.03&  17982&  150 \\        
\textit{033} & \mathrm{B}           &         13\ 14\ 10.21,+51\ 59\ 34.0      & 15.34&   9875&  150 \\        
         034 & \mathrm{B}           &         13\ 14\ 11.77,+52\ 04\ 28.2      & 19.49&  83207&   36 \\
         035 & \mathrm{B}           &         13\ 14\ 14.92,+51\ 59\ 20.6      & 20.54&  81654&   48 \\
         036 & \mathrm{B}           &         13\ 14\ 15.76,+51\ 47\ 25.5      & 20.39&  84610&   42 \\
\textit{037} & \mathrm{B}           &         13\ 14\ 16.22,+51\ 39\ 20.6      & 21.10& 151044&   45 \\
\textit{038} & \mathrm{B}           &         13\ 14\ 16.99,+51\ 41\ 18.2      & 17.37&  30336&  150 \\        
         039 & \mathrm{B}           &         13\ 14\ 17.60,+51\ 50\ 45.7      & 20.69&  84365&   42 \\
\textit{040} & \mathrm{B}           &         13\ 14\ 17.76,+51\ 57\ 03.2      & 15.98&  17586&  150 \\        
\textit{041} & \mathrm{B}           &         13\ 14\ 17.93,+51\ 59\ 45.5      & 18.29& 230894&  150 \\        
\textit{042} & \mathrm{B}           &         13\ 14\ 20.11,+51\ 59\ 09.1      & 20.87& 103548&   63 \\
                        \noalign{\smallskip}			    
            \hline					    
            \noalign{\smallskip}			    
            \hline					    
         \end{array}
     $$ 
         \end{table}


%% file: tabdedica2dz.tex
\begin{table}
        \caption[]{Substructures detected by the analysis of the 2D distribution of the spectroscopic members of A1703. For each subclump, the 2D-DEDICA method provides the number of assigned member galaxies $N_{\rm S}$, right ascension and declination of the density peak, the relative density with respect to the densest subclump $\rho_{\rm S}$, and the $\chi^2$ value of the galaxy peak.}
         \label{tabdedica2dz}
            $$
         \begin{array}{c r c c r }
            \hline
            \noalign{\smallskip}
            \hline
            \noalign{\smallskip}
\mathrm{Subclump} & N_{\rm S} & \alpha({\rm J}2000),\,\delta({\rm J}2000)&\rho_{
\rm S}&\chi^2_{\rm S}\\
& & \mathrm{h:m:s,\degree:\arcmm:\arcs}&&\\
         \hline
         \noalign{\smallskip}
\mathrm{N}      & 60&13\ 15\ 03.8,+51\ 49\ 50&1.00&33.9\\
\mathrm{S}      & 59&13\ 15\ 06.1,+51\ 49\ 00&0.95&27.4\\
              \noalign{\smallskip}
              \noalign{\smallskip}
            \hline
            \noalign{\smallskip}
            \hline
         \end{array}
$$
\end{table}

%% file: tabhtree.tex
\begin{table}
        \caption[]{Substructures detected by the 3D analysis of the spectroscopic members of A1703. For each subclump, the Htree method provides the number of assigned member galaxies $N_{\rm S}$, the mean radial velocity with its error, and the radial velocity dispersion with its error.}
         \label{tabhtree}
            $$
         \begin{array}{l r c c}
            \hline
            \noalign{\smallskip}
            \hline
            \noalign{\smallskip}
\mathrm{Subclump} & N_{\rm S} & \left<V\right>& \sigma_{\rm v} \\
 & &\mathrm{km}\,\,\mathrm{s}^{-1}&\mathrm{km}\,\,\mathrm{s}^{-1} \\
         \hline
         \noalign{\smallskip}
\mathrm{G1}      & 47& 82022\pm 85& 577\pm 48\\
\mathrm{G2}      & 35& 84681\pm 99& 578\pm 75\\
\mathrm{G11}     & 21& 82466\pm 83& 370\pm 69\\
\mathrm{G12}     & 12& 81242\pm 73& 236\pm 72\\
              \noalign{\smallskip}
              \noalign{\smallskip}
            \hline
            \noalign{\smallskip}
            \hline
         \end{array}
$$
\end{table}

%% file: tabkmm3d.tex
\begin{table}
        \caption[]{Substructures detected by the 3D analysis of the spectroscopic members of A1703. For each subclump, the 3D-KMM method provides the number of assigned member galaxies $N_{\rm S}$, the mean radial velocity with its error, and the radial velocity dispersion with its error.}
         \label{tabkmm3d}
            $$
         \begin{array}{l r c c}
            \hline
            \noalign{\smallskip}
            \hline
            \noalign{\smallskip}
\mathrm{Subclump} & N_{\rm S} & \left<V\right>& \sigma_{\rm v} \\
 & &\mathrm{km}\,\,\mathrm{s}^{-1}&\mathrm{km}\,\,\mathrm{s}^{-1} \\
         \hline
         \noalign{\smallskip}
\mathrm{KMM3D1} & 86& 82119\pm \,\,\,\,85& 782\pm 66\\
\mathrm{KMM3D2} & 61& 84608\pm 109& 845\pm 97\\
              \noalign{\smallskip}
              \noalign{\smallskip}
            \hline
            \noalign{\smallskip}
            \hline
         \end{array}
$$
\end{table}

%% file: tabdedica3dz.tex
\begin{table}
        \caption[]{Substructures detected by the 3D analysis of the spectroscopic members of A1703. For each subclump, the 3D-DEDICA method provides the number of assigned member galaxies $N_{\rm S}$, the fitted radial velocity $V_{\rm fit}$, right ascension and declination of the density peak, the relative density with respect to the densest subclump $\rho_{\rm S}$, and the $\chi^2$ value of the galaxy peak.}
         \label{tabdedica3dz}
            $$
         \begin{array}{l r c c c r }
            \hline
            \noalign{\smallskip}
            \hline
            \noalign{\smallskip}
\mathrm{Subclump} & N_{\rm S} & V_{\rm fit} &\alpha({\rm J}2000),\,\delta({\rm J}2000)&\rho_{\rm S}&\chi^2_{\rm S}\\
& &\mathrm{km}\,\,\mathrm{s}^{-1} & \mathrm{h:m:s,\degree:\arcmm:\arcs}&&\\
         \hline
         \noalign{\smallskip}
\mathrm{DED3D1}      & 49& 84346& 13\ 15\ 04.9,+51\ 49\ 12&0.40&30\\
\mathrm{DED3D2}      & 43& 81242& 13\ 15\ 02.8,+51\ 50\ 16&1.00&44\\
\mathrm{DED3D3}      & 26& 82771& 13\ 15\ 07.9,+51\ 48\ 36&0.32&20\\
              \noalign{\smallskip}
              \noalign{\smallskip}
            \hline
            \noalign{\smallskip}
            \hline
         \end{array}
$$
\end{table}

%% file: catalogA1703a1b.tex

\addtocounter{table}{-5}
\begin{table}
          \caption[ ]{Continued.}
     $$ 
           \begin{array}{r c c c r r}
            \hline
            \noalign{\smallskip}
            \hline
            \noalign{\smallskip}

\mathrm{ID} & \mathrm{Source} & \mathrm{\alpha},\mathrm{\delta}\,(\mathrm{J}2000)  & r & V\,\,\,\,\,&\mathrm{\Delta}V\\
  & &                 & &\mathrm{(\,km}&\mathrm{s^{-1}\,)}\\

            \hline
            \noalign{\smallskip}

         043 & \mathrm{B}           &         13\ 14\ 21.80,+51\ 54\ 28.3      & 19.60&  82689&   27 \\
         044 & \mathrm{B}           &         13\ 14\ 22.78,+51\ 48\ 55.1      & 20.03&  83906&   42 \\
\textit{045} & \mathrm{B}           &         13\ 14\ 23.69,+51\ 41\ 55.0      & 17.35&  30183&  150 \\      
         046 & \mathrm{B}           &         13\ 14\ 26.19,+51\ 50\ 20.3      & 19.25&  82893&   24 \\
\textit{047} & \mathrm{B}           &         13\ 14\ 27.08,+52\ 01\ 14.4      & 19.79& 106429&   24 \\
\textit{048} & \mathrm{B}           &         13\ 14\ 32.96,+52\ 02\ 12.2      & 18.98& 759027&  150 \\     
         049 & \mathrm{B}           &         13\ 14\ 33.36,+51\ 55\ 04.2      & 20.39&  82611&   51 \\
         050 & \mathrm{B}           &         13\ 14\ 34.01,+51\ 48\ 11.9      & 19.85&  82707&   30 \\
\textit{051} & \mathrm{B}           &         13\ 14\ 34.53,+51\ 39\ 47.5      & 17.56&  53363&  150 \\     
\textit{052} & \mathrm{B}           &         13\ 14\ 35.86,+52\ 00\ 11.3      & 16.88&  25995&  150 \\     
         053 & \mathrm{B}           &         13\ 14\ 38.16,+51\ 44\ 43.7      & 21.72&  84428&   69 \\
         054 & \mathrm{B}           &         13\ 14\ 38.65,+51\ 56\ 10.3      & 20.17&  82281&   54 \\
\textit{055} & \mathrm{B}           &         13\ 14\ 38.71,+51\ 44\ 07.6      & 19.56&  92369&   27 \\
\textit{056} & \mathrm{L}           &         13\ 14\ 41.80,+51\ 48\ 57.5      & 17.12&  56244&   10 \\
         057 & \mathrm{B}           &         13\ 14\ 42.01,+51\ 58\ 42.5      & 21.28&  83045&   72 \\
\textit{058} & \mathrm{L}           &         13\ 14\ 42.92,+51\ 48\ 16.3      & 19.95& 143644&   34 \\
         059 & \mathrm{B}           &         13\ 14\ 44.28,+51\ 51\ 03.7      & 20.34&  84679&   48 \\
\textit{060} & \mathrm{B}           &         13\ 14\ 45.30,+51\ 55\ 15.2      & 19.64&  80920&   30 \\
\textit{061} & \mathrm{B}           &         13\ 14\ 45.60,+51\ 38\ 52.8      & 21.01& 102172&  108 \\
         062 & \mathrm{T}           &         13\ 14\ 45.78,+51\ 51\ 05.3      & 19.96&  85369&  130 \\
         063 & \mathrm{B}           &         13\ 14\ 46.37,+51\ 47\ 52.0      & 20.15&  84134&   39 \\
\textit{064} & \mathrm{B}           &         13\ 14\ 47.76,+51\ 43\ 43.1      & 21.59& 143346&   54 \\
         065 & \mathrm{B}           &         13\ 14\ 48.48,+51\ 52\ 46.5      & 20.90&  82956&   69 \\
\textit{066} & \mathrm{B}           &         13\ 14\ 48.94,+52\ 03\ 24.4      & 17.77&  90741&  150 \\     
\textit{067} & \mathrm{B}           &         13\ 14\ 49.29,+51\ 40\ 48.0      & 17.63&  29904&  150 \\     
\textit{068} & \mathrm{B}           &         13\ 14\ 49.50,+52\ 02\ 23.0      & 16.78&  52874&  150 \\     
         069 & \mathrm{B}           &         13\ 14\ 50.39,+51\ 58\ 46.2      & 20.58&  82485&   42 \\
         070 & \mathrm{B}           &         13\ 14\ 50.66,+52\ 00\ 58.7      & 21.08&  82488&   72 \\
         071 & \mathrm{T}           &         13\ 14\ 51.28,+51\ 50\ 07.2      & 20.29&  83357&   77 \\
\textit{072} & \mathrm{L}           &         13\ 14\ 51.28,+51\ 50\ 11.3      & 20.41& 168269&   38 \\
         073 & \mathrm{T}           &         13\ 14\ 51.41,+51\ 49\ 41.3      & 19.28&  81636&   50 \\
         074 & \mathrm{B}           &         13\ 14\ 52.38,+51\ 48\ 14.6      & 20.60&  85846&   66 \\
\textit{075} & \mathrm{B}           &         13\ 14\ 52.39,+52\ 02\ 20.8      & 19.17& 102700&   18 \\
         076 & \mathrm{B}           &         13\ 14\ 52.73,+51\ 40\ 31.2      & 18.23&  85060&   21 \\
         077 & \mathrm{B}           &         13\ 14\ 52.84,+51\ 52\ 41.4      & 20.34&  83657&   51 \\
\textit{078} & \mathrm{T}           &         13\ 14\ 53.56,+51\ 52\ 19.4      & 19.79&  52842&   62 \\
\textit{079} & \mathrm{T}           &         13\ 14\ 54.53,+51\ 52\ 21.5      & 20.72&  17931&  145 \\
         080 & \mathrm{T}           &         13\ 14\ 54.68,+51\ 50\ 55.8      & 19.58&  81561&   58 \\
\textit{081} & \mathrm{L}           &         13\ 14\ 54.73,+51\ 53\ 16.9      & 16.85&  53548&   10 \\
         082 & \mathrm{T}           &         13\ 14\ 55.33,+51\ 47\ 53.0      & 18.79&  84662&   75 \\
\textit{083} & \mathrm{L}           &         13\ 14\ 55.38,+51\ 51\ 46.9      & 21.25& 184791&  104 \\
\textit{084} & \mathrm{B}           &         13\ 14\ 55.51,+51\ 41\ 03.0      & 19.65& 576687&  150 \\

                        \noalign{\smallskip}			    
            \hline					    
            \noalign{\smallskip}			    
            \hline					    
         \end{array}
     $$ 
         \end{table}
\addtocounter{table}{-1}
\begin{table}
          \caption[ ]{Continued.}
     $$ 
           \begin{array}{r c c c r r}
            \hline
            \noalign{\smallskip}
            \hline
            \noalign{\smallskip}

\mathrm{ID} & \mathrm{Source} & \mathrm{\alpha},\mathrm{\delta}\,(\mathrm{J}2000)  & r & V\,\,\,\,\,&\mathrm{\Delta}V\\
  & &                 & &\mathrm{(\,km}&\mathrm{s^{-1}\,)}\\

            \hline
            \noalign{\smallskip}

\textit{085} & \mathrm{B}           &         13\ 14\ 55.52,+52\ 03\ 45.2      & 20.44&  90810&   51 \\ 
         086 & \mathrm{B}           &         13\ 14\ 55.69,+51\ 47\ 19.0      & 19.91&  81514&   30 \\
\textit{087} & \mathrm{T}           &         13\ 14\ 55.79,+51\ 53\ 02.6      & 20.32&  53080&  100 \\
\textit{088} & \mathrm{B}           &         13\ 14\ 55.85,+51\ 54\ 28.4      & 20.72&  80851&   66 \\
\textit{089} & \mathrm{T}           &         13\ 14\ 56.12,+51\ 52\ 57.2      & 19.38&  53618&   90 \\
         090 & \mathrm{B}           &         13\ 14\ 56.25,+51\ 53\ 09.9      & 20.99&  81151&   60 \\
         091 & \mathrm{T}           &         13\ 14\ 56.40,+51\ 50\ 52.4      & 20.87&  80925&  122 \\
\textit{092} & \mathrm{L}           &         13\ 14\ 56.83,+51\ 48\ 53.0      & 20.79& 733766&  263 \\
         093 & \mathrm{T}           &         13\ 14\ 56.90,+51\ 51\ 00.5      & 19.84&  82293&   77 \\
         094 & \mathrm{T}           &         13\ 14\ 57.88,+51\ 51\ 41.8      & 20.46&  81642&  115 \\
         095 & \mathrm{B}           &         13\ 14\ 58.09,+51\ 49\ 16.2      & 21.19&  86520&  180 \\
\textit{096} & \mathrm{T}           &         13\ 14\ 58.76,+51\ 52\ 14.0      & 19.39&  53872&   41 \\
         097 & \mathrm{B}           &         13\ 14\ 58.77,+51\ 44\ 22.6      & 20.42&  83594&   33 \\
         098 & \mathrm{T}           &         13\ 14\ 58.77,+51\ 49\ 50.9      & 19.30&  84116&   70 \\
\textit{099} & \mathrm{T}           &         13\ 14\ 59.06,+51\ 53\ 13.5      & 18.20&  52547&   39 \\
         100 & \mathrm{T}           &         13\ 14\ 59.15,+51\ 51\ 38.4      & 20.15&  81889&   72 \\
         101 & \mathrm{T}           &         13\ 14\ 59.32,+51\ 49\ 46.6      & 20.01&  80833&  115 \\
         102 & \mathrm{T}           &         13\ 14\ 59.54,+51\ 49\ 57.8      & 21.11&  83505&  137 \\
         103 & \mathrm{B}           &         13\ 14\ 59.56,+51\ 47\ 52.3      & 21.06&  82725&   63 \\
\textit{104} & \mathrm{B}           &         13\ 14\ 59.56,+52\ 02\ 44.4      & 19.70& 102763&   27 \\
         105 & \mathrm{T}           &         13\ 14\ 59.91,+51\ 50\ 10.7      & 20.59&  85323&  125 \\
         106 & \mathrm{T}           &         13\ 15\ 00.12,+51\ 49\ 05.7      & 20.00&  83368&   90 \\
         107 & \mathrm{T}           &         13\ 15\ 00.12,+51\ 50\ 30.5      & 19.89&  81193&   97 \\
         108 & \mathrm{T}           &         13\ 15\ 00.56,+51\ 50\ 03.0      & 20.84&  86270&  190 \\
         109 & \mathrm{T}           &         13\ 15\ 02.01,+51\ 51\ 06.5      & 19.72&  81248&   53 \\
         110 & \mathrm{T}           &         13\ 15\ 02.02,+51\ 47\ 47.7      & 20.40&  84399&  102 \\
\textit{111} & \mathrm{B}           &         13\ 15\ 02.22,+51\ 42\ 37.4      & 19.99&  90645&   45 \\
         112 & \mathrm{B}           &         13\ 15\ 02.25,+51\ 49\ 51.3      & 21.27&  81154&  180 \\
         113 & \mathrm{T}           &         13\ 15\ 02.41,+51\ 50\ 19.3      & 19.09&  81157&   65 \\
         114 & \mathrm{T}           &         13\ 15\ 02.51,+51\ 52\ 19.6      & 21.10&  84816&  135 \\
         115 & \mathrm{T}           &         13\ 15\ 02.87,+51\ 51\ 48.3      & 20.50&  81548&  145 \\
         116 & \mathrm{T}           &         13\ 15\ 02.89,+51\ 49\ 43.8      & 20.41&  84247&  105 \\
         117 & \mathrm{T}           &         13\ 15\ 03.02,+51\ 47\ 02.8      & 20.07&  84860&  100e\\
         118 & \mathrm{T}           &         13\ 15\ 03.03,+51\ 48\ 50.5      & 20.76&  86609&  117 \\
         119 & \mathrm{T}           &         13\ 15\ 03.14,+51\ 46\ 31.5      & 19.55&  84175&   55 \\
         120 & \mathrm{T}           &         13\ 15\ 03.19,+51\ 49\ 56.7      & 18.76&  81117&   67 \\
         121 & \mathrm{T}           &         13\ 15\ 03.32,+51\ 50\ 22.2      & 21.26&  82004&  117 \\
         122 & \mathrm{T}           &         13\ 15\ 03.33,+51\ 50\ 29.7      & 20.58&  81617&   56 \\
         123 & \mathrm{T}           &         13\ 15\ 03.39,+51\ 51\ 33.6      & 20.20&  80986&   97 \\
         124 & \mathrm{T}           &         13\ 15\ 03.57,+51\ 52\ 38.1      & 18.07&  82492&   36 \\
         125 & \mathrm{T}           &         13\ 15\ 03.90,+51\ 49\ 26.8      & 19.61&  84224&   57 \\
         126 & \mathrm{T}           &         13\ 15\ 03.92,+51\ 49\ 09.9      & 20.51&  80792&  120 \\

                        \noalign{\smallskip}			    
            \hline					    
            \noalign{\smallskip}			    
            \hline					    
         \end{array}
     $$ 
\end{table}
\addtocounter{table}{-1}
\begin{table}
          \caption[ ]{Continued.}
     $$ 
           \begin{array}{r c c c r r}
            \hline
            \noalign{\smallskip}
            \hline
            \noalign{\smallskip}

\mathrm{ID} & \mathrm{Source} & \mathrm{\alpha},\mathrm{\delta}\,(\mathrm{J}2000)  & r & V\,\,\,\,\,&\mathrm{\Delta}V\\
  & &                 & &\mathrm{(\,km}&\mathrm{s^{-1}\,)}\\

            \hline
            \noalign{\smallskip}

         127 & \mathrm{T}           &         13\ 15\ 03.92,+51\ 50\ 44.8      & 19.92&  81334&   82 \\
         128 & \mathrm{B}           &         13\ 15\ 04.05,+51\ 39\ 51.0      & 20.35&  84110&   48 \\
         129 & \mathrm{T}           &         13\ 15\ 04.20,+51\ 50\ 04.4      & 20.26&  80779&  100 \\
         130 & \mathrm{B}           &         13\ 15\ 04.26,+51\ 47\ 50.8      & 23.42&  83942&  300 \\
         131 & \mathrm{T}           &         13\ 15\ 04.31,+51\ 47\ 22.6      & 20.55&  83922&   63 \\
\textit{132} & \mathrm{T}           &         13\ 15\ 04.33,+51\ 51\ 33.7      & 18.48&  53157&   75e\\
         133 & \mathrm{T}           &         13\ 15\ 04.35,+51\ 51\ 47.1      & 20.66&  82945&   92 \\
         134 & \mathrm{T}           &         13\ 15\ 04.50,+51\ 49\ 11.1      & 21.57&  84697&  145 \\
         135 & \mathrm{T}           &         13\ 15\ 04.84,+51\ 49\ 32.0      & 18.74&  81410&   65 \\
         136 & \mathrm{T}           &         13\ 15\ 05.00,+51\ 50\ 02.7      & 19.33&  83533&   72 \\
         137 & \mathrm{B}           &         13\ 15\ 05.04,+51\ 46\ 06.3      & 22.58&  87210&  150 \\
         138 & \mathrm{T}           &         13\ 15\ 05.04,+51\ 52\ 08.2      & 20.27&  81517&  105 \\
         139 & \mathrm{T}           &         13\ 15\ 05.10,+51\ 49\ 54.0      & 20.50&  81889&  102 \\
         140 & \mathrm{B}           &         13\ 15\ 05.18,+52\ 03\ 06.9      & 20.05&  81331&   30 \\
\textbf{141} & \mathrm{T}           &         13\ 15\ 05.24,+51\ 49\ 02.6      & 16.76&  84697&   75 \\
         142 & \mathrm{T}           &         13\ 15\ 05.28,+51\ 48\ 35.3      & 20.92&  83155&   80 \\
         143 & \mathrm{T}           &         13\ 15\ 05.29,+51\ 47\ 53.1      & 18.37&  79194&   32 \\
         144 & \mathrm{B}           &         13\ 15\ 05.32,+51\ 45\ 36.5      & 22.04&  82923&  150 \\
         145 & \mathrm{T}           &         13\ 15\ 05.50,+51\ 50\ 01.7      & 20.46&  81449&   72 \\
         146 & \mathrm{T}           &         13\ 15\ 05.75,+51\ 52\ 31.7      & 19.15&  83298&   54 \\
         147 & \mathrm{T}           &         13\ 15\ 05.92,+51\ 48\ 45.1      & 19.16&  82249&   65 \\
         148 & \mathrm{T}           &         13\ 15\ 06.02,+51\ 49\ 01.4      & 21.22&  79868&   97 \\
         149 & \mathrm{T}           &         13\ 15\ 06.17,+51\ 49\ 43.0      & 20.60&  83816&   75 \\
         150 & \mathrm{T}           &         13\ 15\ 06.24,+51\ 48\ 34.9      & 18.34&  82684&  117 \\
         151 & \mathrm{L}           &         13\ 15\ 06.36,+51\ 54\ 27.9      & 18.13&  82178&   12 \\
\textit{152} & \mathrm{T}           &         13\ 15\ 06.52,+51\ 51\ 00.5      & 19.72&  26235&  100e\\
         153 & \mathrm{T}           &         13\ 15\ 06.68,+51\ 48\ 00.3      & 20.08&  85795&   75 \\
\textit{154} & \mathrm{T}           &         13\ 15\ 06.69,+51\ 47\ 16.3      & 19.69& 116336&  288e\\
         155 & \mathrm{T}           &         13\ 15\ 06.96,+51\ 49\ 25.8      & 19.69&  85901&   72 \\
         156 & \mathrm{T}           &         13\ 15\ 07.05,+51\ 50\ 49.9      & 20.46&  83602&  100 \\
         157 & \mathrm{T}           &         13\ 15\ 07.27,+51\ 51\ 20.6      & 20.07&  84067&   80 \\
         158 & \mathrm{T}           &         13\ 15\ 07.41,+51\ 48\ 22.6      & 19.05&  85381&   37 \\
         159 & \mathrm{T}           &         13\ 15\ 07.43,+51\ 48\ 37.9      & 21.34&  83013&  102 \\
         160 & \mathrm{B}           &         13\ 15\ 07.45,+51\ 39\ 35.6      & 20.00&  84730&   51 \\
         161 & \mathrm{T}           &         13\ 15\ 07.47,+51\ 47\ 36.2      & 20.81&  83184&  105 \\
\textit{162} & \mathrm{T}           &         13\ 15\ 07.71,+51\ 51\ 03.4      & 17.80&  25921&  100e\\
         163 & \mathrm{T}           &         13\ 15\ 07.83,+51\ 48\ 58.0      & 18.55&  84172&   62 \\
         164 & \mathrm{T}           &         13\ 15\ 08.08,+51\ 48\ 48.6      & 19.19&  84643&   82 \\
         165 & \mathrm{T}           &         13\ 15\ 08.08,+51\ 53\ 08.0      & 18.93&  85814&   49 \\
         166 & \mathrm{T}           &         13\ 15\ 08.11,+51\ 52\ 17.0      & 20.64&  81998&   97 \\
         167 & \mathrm{T}           &         13\ 15\ 08.19,+51\ 51\ 53.9      & 18.88&  82376&   42 \\
         168 & \mathrm{T}           &         13\ 15\ 08.21,+51\ 48\ 28.9      & 20.78&  87478&  106 \\

                        \noalign{\smallskip}			    
            \hline					    
            \noalign{\smallskip}			    
            \hline					    
         \end{array}
     $$ 
\end{table}
\addtocounter{table}{-1}
\begin{table}
          \caption[ ]{Continued.}
     $$ 
           \begin{array}{r c c c r r}
            \hline
            \noalign{\smallskip}
            \hline
            \noalign{\smallskip}

\mathrm{ID} & \mathrm{Source} & \mathrm{\alpha},\mathrm{\delta}\,(\mathrm{J}2000)  & r & V\,\,\,\,\,&\mathrm{\Delta}V\\
  & &                 & &\mathrm{(\,km}&\mathrm{s^{-1}\,)}\\

            \hline
            \noalign{\smallskip}

         169 & \mathrm{B}           &         13\ 15\ 08.83,+51\ 45\ 45.5      & 23.01&  83372&  150 \\
         170 & \mathrm{B}           &         13\ 15\ 08.91,+51\ 46\ 22.4      & 22.31&  81094&  150 \\
         171 & \mathrm{T}           &         13\ 15\ 08.91,+51\ 49\ 06.7      & 20.99&  82229&  180 \\
         172 & \mathrm{T}           &         13\ 15\ 09.01,+51\ 49\ 55.8      & 20.43&  87915&  112 \\
         173 & \mathrm{T}           &         13\ 15\ 09.40,+51\ 47\ 21.7      & 19.78&  84361&   55 \\
         174 & \mathrm{T}           &         13\ 15\ 09.76,+51\ 48\ 21.5      & 20.74&  81446&  107e\\
\textit{175} & \mathrm{T}           &         13\ 15\ 09.97,+51\ 53\ 31.5      & 21.04& 184767&   60 \\
\textit{176} & \mathrm{T}           &         13\ 15\ 10.01,+51\ 52\ 43.6      & 20.31& 124393&  165 \\
         177 & \mathrm{B}           &         13\ 15\ 10.05,+51\ 43\ 13.3      & 20.20&  83252&   45 \\
\textit{178} & \mathrm{T}           &         13\ 15\ 10.11,+51\ 52\ 23.2      & 19.87&  93275&  107 \\
         179 & \mathrm{T}           &         13\ 15\ 10.15,+51\ 50\ 03.2      & 19.31&  82684&  117 \\
\textit{180} & \mathrm{T}           &         13\ 15\ 10.18,+51\ 53\ 20.0      & 19.93&  52960&  100e\\
\textit{181} & \mathrm{T}           &         13\ 15\ 11.06,+51\ 49\ 03.2      & 17.82&  52937&   62 \\
         182 & \mathrm{B}           &         13\ 15\ 11.06,+51\ 46\ 53.8      & 21.97&  80644&  180 \\
         183 & \mathrm{B}           &         13\ 15\ 11.07,+51\ 45\ 57.7      & 21.36&  82923&  150 \\
         184 & \mathrm{B}           &         13\ 15\ 11.19,+51\ 48\ 23.3      & 20.73&  85069&   63 \\
         185 & \mathrm{B}           &         13\ 15\ 11.20,+51\ 44\ 18.2      & 20.53&  84958&   69 \\
         186 & \mathrm{T}           &         13\ 15\ 11.50,+51\ 46\ 41.5      & 19.77&  80732&   54 \\
         187 & \mathrm{T}           &         13\ 15\ 11.59,+51\ 49\ 20.4      & 19.04&  80514&   55 \\
         188 & \mathrm{B}           &         13\ 15\ 11.96,+51\ 45\ 53.0      & 20.55&  81891&   54 \\
         189 & \mathrm{T}           &         13\ 15\ 12.03,+51\ 47\ 22.4      & 19.55&  81821&   72 \\
         190 & \mathrm{T}           &         13\ 15\ 13.47,+51\ 50\ 04.9      & 19.04&  86158&   57 \\
         191 & \mathrm{T}           &         13\ 15\ 13.56,+51\ 49\ 49.2      & 19.34&  82611&   65 \\
         192 & \mathrm{T}           &         13\ 15\ 13.73,+51\ 49\ 39.5      & 20.39&  85308&   80 \\
         193 & \mathrm{T}           &         13\ 15\ 13.73,+51\ 49\ 55.1      & 20.18&  83418&  125 \\
\textit{194} & \mathrm{T}           &         13\ 15\ 13.73,+51\ 51\ 50.1      & 20.91& 795600&  335e\\
         195 & \mathrm{T}           &         13\ 15\ 13.98,+51\ 49\ 10.8      & 19.75&  80841&   90 \\
         196 & \mathrm{T}           &         13\ 15\ 14.53,+51\ 50\ 03.3      & 19.63&  81590&   50 \\
         197 & \mathrm{T}           &         13\ 15\ 14.69,+51\ 48\ 17.7      & 20.18&  82642&  100 \\
         198 & \mathrm{T}           &         13\ 15\ 14.72,+51\ 47\ 20.1      & 20.43&  82698&  102 \\
         199 & \mathrm{T}           &         13\ 15\ 14.81,+51\ 48\ 30.1      & 20.36&  79793&   72 \\
         200 & \mathrm{T}           &         13\ 15\ 15.18,+51\ 48\ 04.1      & 18.90&  82526&   62 \\
         201 & \mathrm{T}           &         13\ 15\ 16.11,+51\ 47\ 46.8      & 19.66&  85026&   65 \\
         202 & \mathrm{T}           &         13\ 15\ 16.29,+51\ 50\ 23.4      & 18.86&  82002&   52 \\
         203 & \mathrm{B}           &         13\ 15\ 17.14,+51\ 57\ 05.6      & 19.90&  84035&   33 \\
\textit{204} & \mathrm{B}           &         13\ 15\ 17.17,+51\ 53\ 57.6      & 19.44&  80425&   24 \\
\textit{205} & \mathrm{B}           &         13\ 15\ 17.62,+51\ 42\ 00.2      & 22.14& 207031&   45 \\
\textit{206} & \mathrm{T}           &         13\ 15\ 17.69,+51\ 47\ 25.6      & 20.32& 116791&   88 \\
         207 & \mathrm{B}           &         13\ 15\ 18.03,+52\ 00\ 55.0      & 19.80&  82938&   30 \\
\textit{208} & \mathrm{B}           &         13\ 15\ 18.38,+51\ 51\ 41.3      & 20.21& 102454&   30 \\
         209 & \mathrm{T}           &         13\ 15\ 18.91,+51\ 48\ 43.7      & 19.48&  85388&   60 \\
         210 & \mathrm{T}           &         13\ 15\ 19.72,+51\ 48\ 08.1      & 19.66&  82082&   70 \\
    
                         \noalign{\smallskip}			    
             \hline					    
            \noalign{\smallskip}			    
            \hline					    
         \end{array}
     $$ 
\end{table}
\addtocounter{table}{-1}
\begin{table}
          \caption[ ]{Continued.}
     $$ 
           \begin{array}{r c c c r r}
            \hline
            \noalign{\smallskip}
            \hline
            \noalign{\smallskip}

\mathrm{ID} & \mathrm{Source} & \mathrm{\alpha},\mathrm{\delta}\,(\mathrm{J}2000)  & r & V\,\,\,\,\,&\mathrm{\Delta}V\\
  & &                 & &\mathrm{(\,km}&\mathrm{s^{-1}\,)}\\

            \hline
            \noalign{\smallskip}

\textit{211} & \mathrm{B}           &         13\ 15\ 19.98,+51\ 38\ 42.0      & 19.81&  91425&   42 \\
         212 & \mathrm{B}           &         13\ 15\ 20.01,+51\ 43\ 40.5      & 21.30&  84344&  102 \\
         213 & \mathrm{B}           &         13\ 15\ 21.87,+51\ 49\ 03.7      & 19.06&  82020&   33 \\
\textit{214} & \mathrm{B}           &         13\ 15\ 23.11,+52\ 01\ 25.3      & 16.85&  17784&  150 \\
\textit{215} & \mathrm{B}           &         13\ 15\ 24.09,+52\ 03\ 59.1      & 15.75&  18020&  150 \\
         216 & \mathrm{T}           &         13\ 15\ 24.22,+51\ 48\ 18.1      & 19.67&  82459&   77 \\
\textit{217} & \mathrm{B}           &         13\ 15\ 24.90,+51\ 54\ 58.9      & 16.80&  34545&  150 \\       
         218 & \mathrm{B}           &         13\ 15\ 25.33,+51\ 51\ 06.1      & 19.12&  86208&   27 \\
         219 & \mathrm{T}           &         13\ 15\ 26.22,+51\ 47\ 55.6      & 19.80&  85366&   82 \\
\textit{220} & \mathrm{B}           &         13\ 15\ 26.50,+51\ 44\ 53.0      & 20.06&  79502&   30 \\
         221 & \mathrm{B}           &         13\ 15\ 26.86,+51\ 43\ 26.6      & 19.66&  84155&   30 \\
\textit{222} & \mathrm{B}           &         13\ 15\ 28.43,+51\ 37\ 52.7      & 21.40& 137335&  111 \\
\textit{223} & \mathrm{B}           &         13\ 15\ 29.91,+51\ 52\ 57.7      & 21.08& 135929&   48 \\
         224 & \mathrm{B}           &         13\ 15\ 29.93,+51\ 55\ 44.0      & 19.58&  84053&   30 \\
         225 & \mathrm{B}           &         13\ 15\ 31.56,+51\ 45\ 34.0      & 21.26&  82521&   54 \\
         226 & \mathrm{B}           &         13\ 15\ 33.48,+51\ 51\ 16.1      & 21.35&  84607&   63 \\
         227 & \mathrm{B}           &         13\ 15\ 33.75,+51\ 58\ 54.8      & 20.39&  81948&   39 \\
\textit{228} & \mathrm{B}           &         13\ 15\ 43.19,+51\ 49\ 50.2      & 20.19& 116071&   54 \\
\textit{229} & \mathrm{B}           &         13\ 15\ 44.27,+51\ 43\ 33.4      & 19.18&  86715&   30 \\
         230 & \mathrm{B}           &         13\ 15\ 45.61,+51\ 52\ 01.9      & 19.19&  82593&   27 \\
\textit{231} & \mathrm{B}           &         13\ 15\ 45.69,+52\ 00\ 28.3      & 22.39& 304781&   93 \\
         232 & \mathrm{B}           &         13\ 15\ 46.20,+51\ 48\ 23.5      & 19.14&  84032&   24 \\
         233 & \mathrm{B}           &         13\ 15\ 46.60,+51\ 51\ 25.2      & 19.36&  82686&   24 \\
\textit{234} & \mathrm{B}           &         13\ 15\ 50.10,+51\ 39\ 16.0      & 20.85& 129262&   57 \\
\textit{235} & \mathrm{B}           &         13\ 15\ 51.52,+51\ 48\ 47.4      & 19.80&  78333&   33 \\
\textit{236} & \mathrm{B}           &         13\ 15\ 52.41,+51\ 44\ 41.9      & 21.29& 159684&   45 \\
\textit{237} & \mathrm{B}           &         13\ 15\ 52.41,+51\ 46\ 41.6      & 20.63& 102907&   51 \\
\textit{238} & \mathrm{B}           &         13\ 15\ 53.47,+51\ 56\ 58.1      & 20.38& 132484&   18 \\
\textit{239} & \mathrm{B}           &         13\ 15\ 55.87,+51\ 53\ 13.3      & 20.10&  56832&   36 \\
\textit{240} & \mathrm{B}           &         13\ 15\ 59.50,+52\ 04\ 14.3      & 20.44&  95229&   45 \\
\textit{241} & \mathrm{B}           &         13\ 16\ 00.79,+51\ 51\ 07.8      & 17.12&  56613&  150 \\     
\textit{242} & \mathrm{B}           &         13\ 16\ 01.28,+51\ 57\ 00.4      & 20.38& 132517&   45 \\
\textit{243} & \mathrm{B}           &         13\ 16\ 04.09,+51\ 43\ 47.0      & 21.54& 117054&   63 \\
\textit{244} & \mathrm{B}           &         13\ 16\ 08.41,+51\ 40\ 25.6      & 20.93&  89371&   75 \\
\textit{245} & \mathrm{B}           &         13\ 16\ 13.15,+52\ 03\ 27.8      & 19.93& 116766&   30 \\
\textit{246} & \mathrm{B}           &         13\ 16\ 13.47,+52\ 04\ 54.6      & 20.95& 100044&   63 \\
\textit{247} & \mathrm{B}           &         13\ 16\ 13.62,+51\ 45\ 08.0      & 21.08& 102475&   57 \\
\textit{248} & \mathrm{B}           &         13\ 16\ 13.79,+51\ 40\ 25.9      & 20.12&  89287&   48 \\
\textit{249} & \mathrm{B}           &         13\ 16\ 15.13,+51\ 38\ 48.8      & 17.37&   7414&  150 \\
\textit{250} & \mathrm{B}           &         13\ 16\ 17.09,+51\ 49\ 02.3      & 15.92&  30006&  150 \\     
\textit{251} & \mathrm{B}           &         13\ 16\ 17.12,+51\ 40\ 34.4      & 18.07&  89866&  150 \\
\textit{252} & \mathrm{B}           &         13\ 16\ 17.61,+51\ 40\ 32.8      & 21.62&  89671&   48 \\

                        \noalign{\smallskip}			    
            \hline					    
            \noalign{\smallskip}			    
            \hline					    
         \end{array}
     $$ 
\end{table}
\addtocounter{table}{-1}
\begin{table}
          \caption[ ]{Continued.}
     $$ 
           \begin{array}{r c c c r r}
            \hline
            \noalign{\smallskip}
            \hline
            \noalign{\smallskip}

\mathrm{ID} & \mathrm{Source} & \mathrm{\alpha},\mathrm{\delta}\,(\mathrm{J}2000)  & r & V\,\,\,\,\,&\mathrm{\Delta}V\\
  & &                  & &\mathrm{(\,km}&\mathrm{s^{-1}\,)}\\

            \hline
            \noalign{\smallskip}

\textit{253} & \mathrm{B}           &         13\ 16\ 19.78,+51\ 37\ 59.5      & 19.42&  96146&   30 \\
\textit{254} & \mathrm{B}           &         13\ 16\ 21.77,+51\ 56\ 56.8      & 20.48&  97516&   54 \\
\textit{255} & \mathrm{B}           &         13\ 16\ 22.55,+52\ 01\ 04.1      & 19.61& 109346&   30 \\
\textit{256} & \mathrm{B}           &         13\ 16\ 24.13,+51\ 49\ 42.9      & 20.99& 142701&   48 \\
\textit{257} & \mathrm{B}           &         13\ 16\ 24.73,+51\ 38\ 19.0      & 20.60& 122696&   18 \\
\textit{258} & \mathrm{B}           &         13\ 16\ 25.12,+51\ 55\ 06.6      & 20.19& 102853&   39 \\
\textit{259} & \mathrm{B}           &         13\ 16\ 26.10,+51\ 50\ 24.7      & 20.34& 100245&   33 \\
\textit{260} & \mathrm{B}           &         13\ 16\ 26.30,+51\ 41\ 28.2      & 19.80& 108648&   36 \\
\textit{261} & \mathrm{B}           &         13\ 16\ 28.11,+52\ 00\ 23.2      & 21.30& 109409&   63 \\
\textit{262} & \mathrm{B}           &         13\ 16\ 29.73,+52\ 03\ 47.4      & 19.42& 117642&   27 \\
\textit{263} & \mathrm{B}           &         13\ 16\ 33.53,+51\ 51\ 53.8      & 21.20& 100386&   78 \\
\textit{264} & \mathrm{B}           &         13\ 16\ 34.26,+52\ 00\ 07.7      & 20.16& 118847&   15 \\
\textit{265} & \mathrm{B}           &         13\ 16\ 35.47,+51\ 54\ 28.7      & 23.11& 185215&   89 \\     
\textit{266} & \mathrm{B}           &         13\ 16\ 37.66,+51\ 55\ 11.8      & 19.93&  90768&   33 \\
         267 & \mathrm{B}           &         13\ 16\ 41.40,+52\ 02\ 21.2      & 19.89&  83510&   39 \\
\textit{268} & \mathrm{B}           &         13\ 16\ 41.45,+51\ 57\ 22.0      & 20.07& 118202&   27 \\
\textit{269} & \mathrm{B}           &         13\ 16\ 43.20,+51\ 42\ 56.8      & 20.89&  88894&   39 \\
\textit{270} & \mathrm{B}           &         13\ 16\ 43.76,+51\ 45\ 54.5      & 20.52& 129894&   30 \\
\textit{271} & \mathrm{B}           &         13\ 16\ 45.52,+51\ 51\ 29.0      & 20.95& 103099&  102 \\
\textit{272} & \mathrm{B}           &         13\ 16\ 46.34,+51\ 40\ 53.8      & 20.81& 101519&   51 \\
\textit{273} & \mathrm{B}           &         13\ 16\ 46.68,+51\ 43\ 34.8      & 21.09& 130098&   48 \\
\textit{274} & \mathrm{B}           &         13\ 16\ 47.00,+51\ 44\ 11.6      & 20.59& 123820&   54 \\
\textit{275} & \mathrm{B}           &         13\ 16\ 48.49,+51\ 56\ 30.5      & 20.73& 117273&   33 \\
\textit{276} & \mathrm{B}           &         13\ 16\ 49.03,+51\ 58\ 13.2      & 18.85&  99969&   33 \\
\textit{277} & \mathrm{B}           &         13\ 16\ 49.96,+51\ 38\ 07.5      & 19.17& 112059&   18 \\
         278 & \mathrm{B}           &         13\ 16\ 51.24,+51\ 38\ 27.4      & 20.59&  83435&   42 \\

                        \noalign{\smallskip}			    
            \hline					    
            \noalign{\smallskip}			    
            \hline					    
         \end{array}
     $$ 
\end{table}
